\documentclass[fleqn,usenatbib]{mnras}

\usepackage{newtxtext,newtxmath}

\usepackage[T1]{fontenc}
\usepackage{ae,aecompl}

\usepackage{graphicx}	
    \graphicspath{{figures/}}
\usepackage{amsmath}	
\usepackage{amssymb}	
\usepackage{rotating}
\usepackage{booktabs}
\usepackage{makecell}

\newcommand{\Solaris}{Solaris}
\newcommand{\TESS}{\textit{TESS}}
\newcommand{\VCentauri}{V1200 Centauri}
\newcommand{\VCen}{V1200~Cen}
\newcommand{\ie}{i.e.\ }
\newcommand{\eg}{e.g.\ }
\newcommand{\JKTEBOP}{\textsc{jktebop}}
\newcommand{\EBOP}{\textsc{ebop}}
\newcommand{\Kepler}{\textit{Kepler}}
\newcommand{\PHOEBE}{\textsc{phoebe}}
\newcommand{\Gaia}{\textit{Gaia}}
\newcommand{\ms}{m~s$^{-1}$}
\newcommand{\IDL}{\textsc{idl}}
\newcommand{\GALUVW}{\textit{gal\_uvw}}
\newcommand{\Hipparcos}{\textit{Hipparcos}}

\title[Analysis of the multiple star \VCentauri{}]{Analysis of eclipsing binaries in multiple stellar systems: the case of \VCentauri{}}

\author[F. Marcadon et al.]{
F. Marcadon,$^{1}$\thanks{E-mail: fmarcadon@ncac.torun.pl}
K. G. He{\l}miniak,$^{1}$
J. P. Marques,$^{2}$
R. Paw{\l}aszek,$^{1}$
P. Sybilski,$^{1}$
\newauthor{
S. K. Koz{\l}owski,$^{1}$
M. Ratajczak$^{3}$
and M. Konacki$^{1}$}
\\
$^{1}$Nicolaus Copernicus Astronomical Center, Polish Academy of Sciences, ul. Rabia\'{n}ska 8, 87-100 Toru\'{n}, Poland\\
$^{2}$Institut d'Astrophysique Spatiale, UMR8617, CNRS, Universit\'{e} Paris-Saclay, B\^{a}timent 121, 91405 Orsay Cedex, France\\
$^{3}$Astronomical Observatory, University of Warsaw, Al. Ujazdowskie 4, 00-478 Warszawa, Poland\\
}

\date{Accepted XXX. Received YYY; in original form ZZZ}

\pubyear{2020}

\begin{document}
\label{firstpage}
\pagerange{\pageref{firstpage}--\pageref{lastpage}}
\maketitle

\begin{abstract}
We present a new analysis of the multiple star \VCentauri{} based on the most recent observations for this system. We used the photometric observations from the \Solaris{} network and the \TESS{} telescope, combined with the new radial velocities from the CHIRON spectrograph and those published in the literature. We confirmed that \VCen{} consists of a 2.5-day eclipsing binary orbited by a third body. We derived the parameters of the eclipsing components, which are $M_{Aa} = 1.393\pm0.018\,$M$_\odot$, $R_{Aa} = 1.407\pm0.014\,$R$_\odot$ and $T_{{\rm eff},Aa} = 6\,588\pm58\,$K for the primary, and $M_{Ab} = 0.863\,3\pm0.008\,1\,$M$_\odot$, $R_{Ab} = 1.154\pm0.014\,$R$_\odot$ and $T_{{\rm eff},Ab} = 4\,475\pm68\,$K for the secondary. Regarding the third body, we obtained significantly different results than previously published. The period of the outer orbit is found to be 180.4 days, implying a minimum mass $M_B = 0.871\pm0.020\,$M$_\odot$. Thus, we argue that \VCen{} is a quadruple system with a secondary pair composed of two low-mass stars. Finally, we determined the ages of each eclipsing component using two evolution codes, namely MESA and CESTAM. We obtained ages of 16--18.5$\,$Myr and 5.5--7$\,$Myr for the primary and the secondary, respectively. In particular, the secondary appears larger and hotter than predicted at the age of the primary. We concluded that dynamical and tidal interactions occurring in multiples may alter the stellar properties and explain the apparent non-coevality of \VCentauri{}.
\end{abstract}




\section{Introduction}

Binary or multiple stellar systems are very common in our Galaxy. Over the past century, stellar duplicity has been reported using different techniques from ground and space-based instruments (\eg spectroscopy, interferometry, photometry). Depending on the nature of the observations, these systems are referred to as spectroscopic, visual, eclipsing or astrometric binaries. More recently, asteroseismology has allowed the discovery of binary stars showing solar-like oscillations in both components of the system, that is, seismic binaries (see \eg \citealt{2018A&A...617A...2M} for a review). For such systems, a model-dependent approach is required to determine their stellar parameters. In this context, eclipsing binaries (EBs) that are also double-lined spectroscopic binaries (SB2s) provide a direct determination of the stellar parameters through their dynamics. The stellar masses and radii can then be measured with exquisite precision below $\sim$1--3$\%$ \citep{2010A&ARv..18...67T}. Precise stellar parameters are actually crucial for calibrating theoretical models of stars, mainly during the pre-main-sequence (PMS) phase, where evolution is more rapid.

Eclipsing binaries with low-mass PMS stars represent a real challenge for theoretical models due to the complexity of the stellar physics involved \citep{2014NewAR..60....1S}. These kind of systems are therefore valuable test cases for models at early stages of stellar evolution. Unfortunately, there are only few systems with such features reported in the literature. \citet{2019A&A...623A..23G} listed 14 known EBs with masses, radii and ages below 1.4$\,{\rm M}_\odot$, 2.4$\,{\rm R}_\odot$ and 17$\,$Myr, respectively (see references therein). Another system was identified as a possible candidate by \citet{2015MNRAS.448.1937C}, namely \VCentauri{}. However, the precision on the derived parameters, in particular stellar radii, did not allow the authors to properly determine the individual ages of the eclipsing components ($\sim$30$\,$Myr). In this work, we propose to re-analyse \VCen{} using the most recent observations of the system. From their radial-velocity analysis, \citet{2015MNRAS.448.1937C} claimed that \VCen{} is a hierarchical triple star system with an outer period of almost one year. Here, we argue that the third body is itself a binary system, making \VCen{} a quadruple star system with a 180-day outer period.

Due to its multiplicity, \VCen{} appears to be an interesting target for studying the dynamical evolution of multiple star systems. In particular, the dynamics of hierarchical quadruple systems is a difficult problem that has been investigated by a number of authors (see \citealt{2019MNRAS.482.2262H}, and references therein). For triple and quadruple systems, it has notably been shown that the period distributions of inner orbits present an enhancement at a few to several tens of days \citep{2008MNRAS.389..925T}. In the case of triple systems, the formation of a short-period binary can be explained by Lidov-Kozai \citep{1962P&SS....9..719L,1962AJ.....67..591K} cycles with tidal friction (see \eg \citealt{2016ComAC...3....6T} for a review). A legitimate question is whether this mechanism extends to quadruple star systems, especially for those with PMS stars in a close orbit. \citet{2019MNRAS.482.2262H} suggested that others processes occurring during the stellar formation may produce such quadruple systems with close inner pairs. It is widely accepted that stars belonging to a binary or multiple system are formed at the same time from the same interstellar material \citep{2002ARA&A..40..349T}. However, this hypothesis needs to be tested in the case of specific systems such as \VCen{}.

This article is organised as follows. Section~\ref{sec:obs} describes the observational data used in this work, including \Solaris{} and \TESS{} photometry as well as radial velocity measurements of \VCen{}. Section~\ref{sec:analysis} presents the light-curve and radial-velocity analysis of the system leading to the determination of the stellar masses and radii for the two eclipsing components. In Section~\ref{sec:discussion}, we discuss the implications of the main features of \VCen{} on its evolutionary status. Finally, the conclusions of this work are summarised in Section~\ref{sec:summary}.
 

\section{Observations}
\label{sec:obs}

\subsection{\Solaris{} photometry}

We collected photometric data for \VCen{} during three main campaigns of observation between 2017 February and August ($\sim$75 nights), between 2018 March and August ($\sim$55 nights) and between 2019 February and April ($\sim$25 nights) with \Solaris{}, a network of  four autonomous observatories in the Southern Hemisphere \citep{2014SPIE.9145E..04K,2017PASP..129j5001K}. The \Solaris{} network aims to detect exoplanets around binaries and multiple stars such as \VCen{} using high cadence and high-precision photometric observations of these systems \citep{2012IAUS..282..111K}. This global network allows a continuous night-time coverage from the end of March until mid-September due to the location of the four stations: \Solaris{}-1 and \Solaris{}-2 in the South African Astronomical Observatory (SAAO)\footnote{\url{https://www.saao.ac.za/}} in South Africa, \Solaris{}-3 in Siding Spring Observatory (SSO)\footnote{\url{https://www.sidingspringobservatory.com.au/}} in Australia and \Solaris{}-4 in Complejo Astron\'{o}mico El Leoncito (CASLEO)\footnote{\url{https://casleo.conicet.gov.ar/}} in Argentina. Each station is equipped with a 0.5-m diameter reflecting telescope. \Solaris{}-3 is a Schmidt-Cassegrain f$/$9 optical system equipped with a field corrector whereas the other telescopes are Ritchey-Chr\'{e}tien f$/$15 optical systems. All four telescopes utilise Andor iKon-L CCD cameras thermoelectrically cooled to $-70^\circ$~Celsius during observations.

\Solaris{} telescopes allow multi-color photometry in ten bands using Johnson ($UBVRI$) and Sloan (u$'$g$'$r$'$i$'$z$'$) filters. Following the work of \citet{2015MNRAS.448.1937C} on \VCen{}, we observed this system both in the $V$ and $I$ bands in order to obtain a better signal-to-noise (S/R) ratio in the infrared (IR) for the secondary eclipse. The image acquisition process of \Solaris{} is described in details in \citeauthor{2017PASP..129j5001K} (\citeyear{2017PASP..129j5001K}; their figure~13) and includes the typical calibration steps, \ie bias and dark frame acquisition as well as flat fielding for all the different filters considered. For the data reduction, we adopted a custom photometric pipeline based on the Photutils\footnote{\url{https://photutils.readthedocs.io/en/stable/}} package of Astropy \citep{Bradley_2019_2533376}. Each raw science image was then bias substrated and corrected for CCD inhomogeneities using dedicated flat-field frames. In this work, we employed the differential photometry method in order to limit the variability of the signal due to atmospheric conditions. As comparison star, we used the closer and brightest target around \VCen{} in the \Solaris{} field of view (13--21 arcmin), namely TYC 7790-1580-1. This latter has a $V$ magnitude of 10.632 \citep{2014AJ....148...81M}. Aperture photometry was performed by defining fixed star apertures and sky background annulus both for the target and the comparison star.

Finally, we applied the W\={o}tan\footnote{\url{https://github.com/hippke/wotan}} detrending algorithm developed by \cite{2019AJ....158..143H} to the light curves obtained from our photometric pipeline. Among the proposed methods, we adopted a sum-of-(co)sines approach associated with an iterative sigma-clipping technique. In order to avoid distortions in the eclipse profiles, W\={o}tan offers the possibility of masking them during detrending. However, at the time of writing, this feature was implemented only for the sum-of-(co)sines approach in W\={o}tan. The final light curves were then obtained by applying the W\={o}tan detrending algorithm with two filters of different widths. Indeed, we took into account both the long-term variability of the comparison star using a 2-day filter and the short-term atmospheric fluctuations using a 2-hour filter. The resulting light curves consist in total of $\sim$30$\,000$ data points in $V$ and $I$, respectively.

\subsection{\TESS{} photometry}

\begin{figure}
    \centering
	\includegraphics[trim = 0.cm 0.0cm 0.0cm 0.0cm,clip,width=0.8\columnwidth,angle=0]{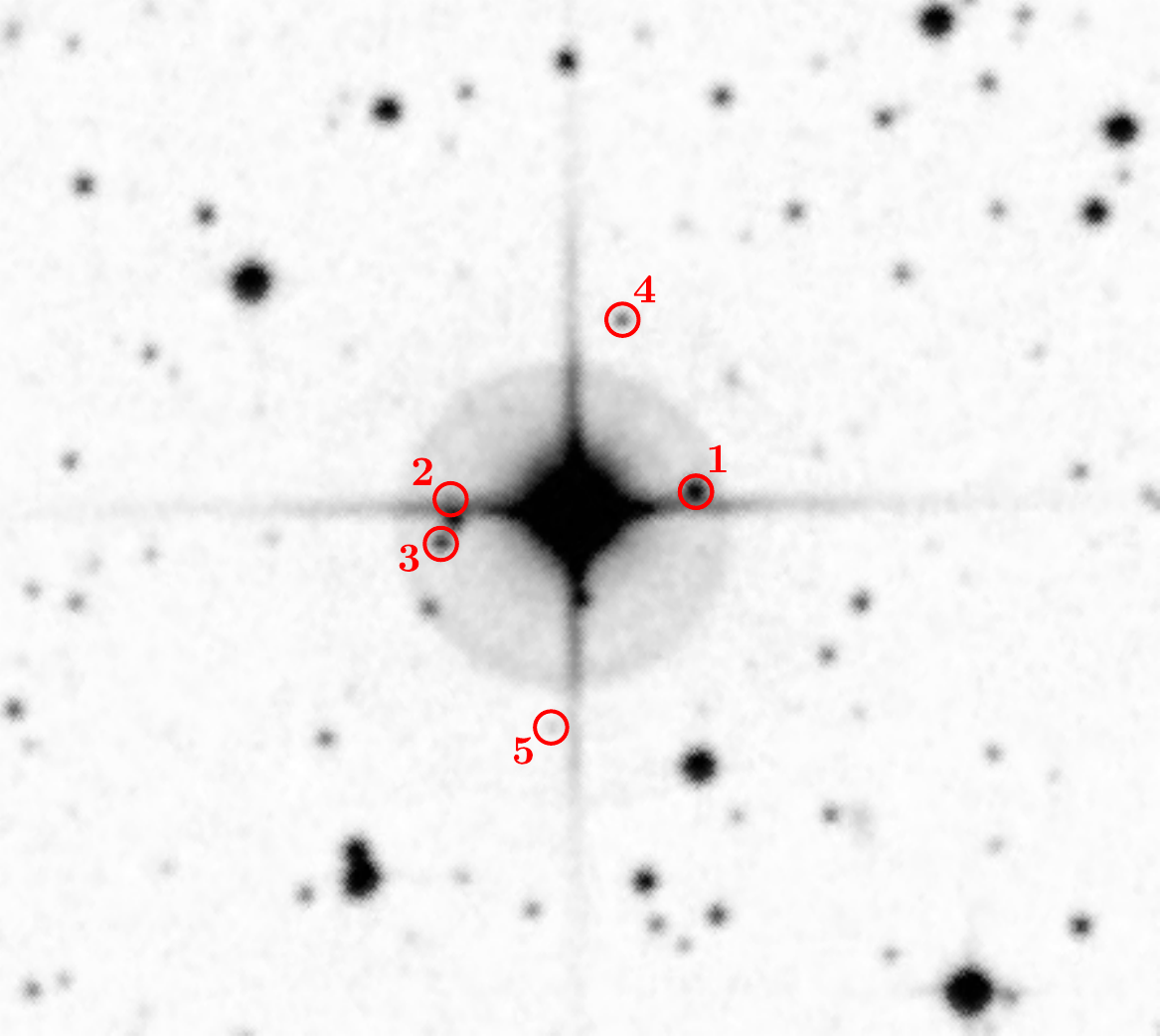}
    \caption{
	A $4 \times 4\,$arcmin$^2$ image centred on \VCen{}. The five stars shown by open red circles are located within the optimal aperture as defined in the Data Validation (DV) Report for \VCen{}. For these five stars, the contribution to the total flux is very small ($<$0.1\%). There are three other stars (not outlined in the picture) that lie within the optimal aperture. However, even taking them into account, the flux excess due to contaminant stars will not be larger than about 0.2--0.3\%.
	}
    \label{fig:DSScol}
\end{figure}

\begin{figure}
	\includegraphics[trim = 0.5cm 0.0cm 1.0cm 1.0cm,clip,width=1.0\columnwidth,angle=0]{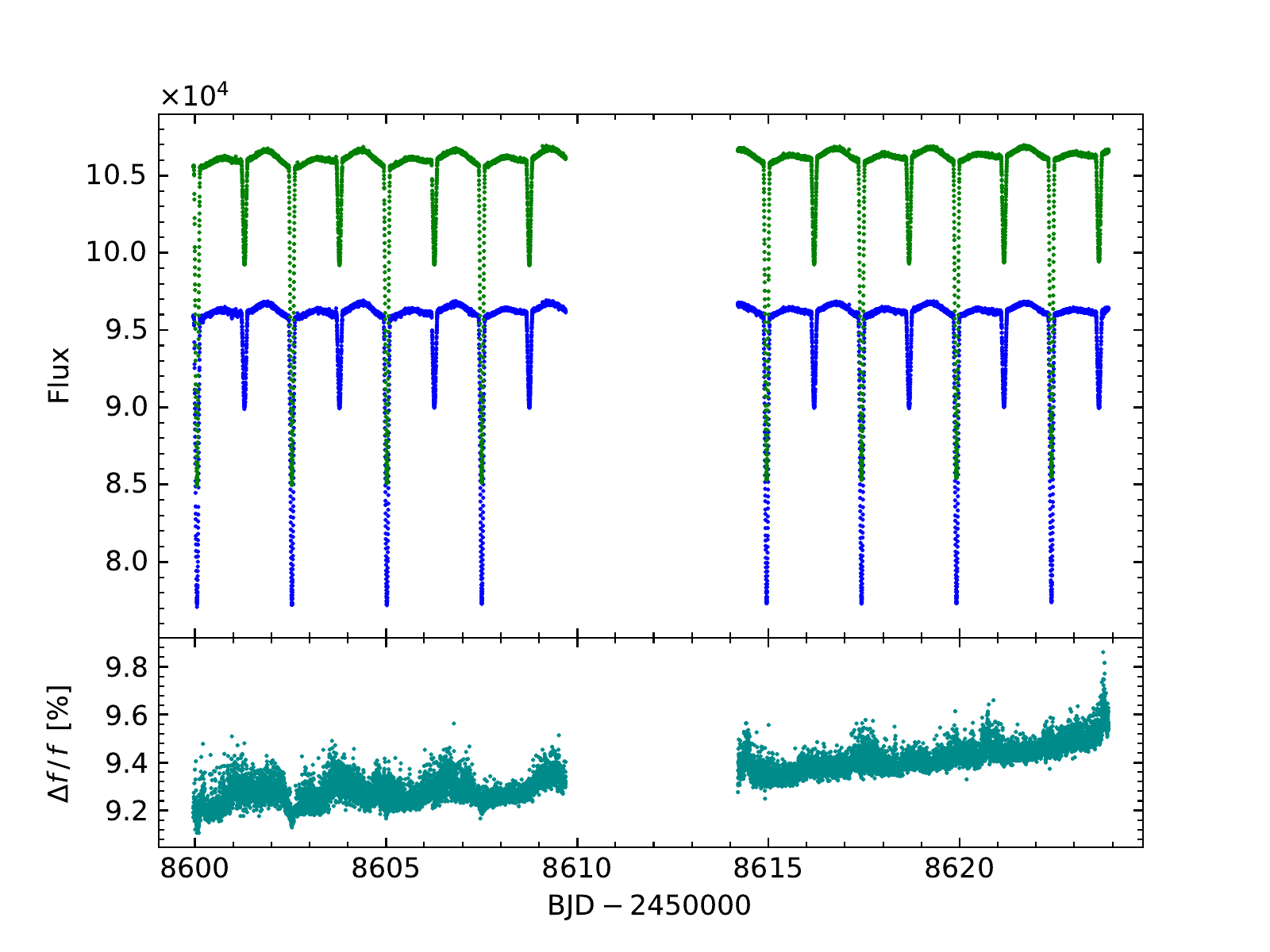}
    \caption{
	SAP (blue) and PDCSAP (green) light curves of \VCen{}. The relative flux difference is plotted in the lower panel. The linear trend seen in the flux difference was taken into account when fitting the SAP and PDCSAP light curves. Furthermore, the variability of the flux difference does not exceed 0.2\%.
	}
    \label{fig:TESS-PDCSAP}
\end{figure}

\VCen{} was observed by the Transiting Exoplanet Survey Satellite (\TESS{}; \citealt{2015JATIS...1a4003R}) in two-minute cadence mode for 27.1~days during sector~11\footnote{Guest Investigator programme G011083, PI: He{\l}miniak}, that is between 2019 April 23 and 2019 May 20. In this work, we analysed the high-precision photometric data collected by \TESS{} for \VCen{} (TIC~166624433), in addition to the \Solaris{} light curves. These data were generated by the \TESS{} Science Processing Operations Center (SPOC; \citealt{2016SPIE.9913E..3EJ}) and made available on the Mikulski Archive for Space Telescopes (MAST)\footnote{\url{https://archive.stsci.edu/}}. In particular, we used the Pre-search Data Conditioning Simple Aperture Photometry (PDCSAP; \citealt{2012PASP..124.1000S,2012PASP..124..985S,2014PASP..126..100S}) version of the light curve, which has been corrected for instrumental effects and contamination by nearby stars. As seen in Fig.~\ref{fig:DSScol}, there are five additional stars identified as \TESS{} targets within the optimal aperture used to extract the light curve of \VCen{}: TIC 166624435, TIC 166624434, TIC 166624431, TIC 166624443, and TIC 166624426. Their corresponding magnitudes in the \TESS{} band are 16.56, 18.18, 17.07, 17.50, and 18.15$\,$mag, respectively. Given that \VCen{} has a \TESS{} magnitude of 7.93$\,$mag, we estimated the contribution of the contaminant stars to the total flux measured by \TESS{} to be lower than 0.1\%. Fig.~\ref{fig:TESS-PDCSAP} shows the comparison between the SAP and PDCSAP light curves of \VCen{}.

The \TESS{} observations of sector~11 correspond to orbits 29 and 30 of the spacecraft around the Earth. At the start of both orbits, camera~1 that observed \VCen{} was disabled due to strong scattered light signals affecting the systematic error removal in PDC\footnote{More details are given in the Data Release note of sector~11 (DR16) available at \url{https://archive.stsci.edu/tess/tess_drn.html}}. A consequence is the presence of two gaps in the time series, which contains $13\,887$ flux measurements, implying a degraded duty cycle of $\sim$71$\%$. Despite this, the orbital period of the eclipsing pair, namely $\sim$2.5~days, is short enough to distinguish a total of 16~eclipse events in the \TESS{} light curve (8~primary and 8~secondary). For each measurement, we converted the flux $f$ into magnitude using the simple relation $m = -2.5\log(f/\Tilde{f})+m_T$, where $\Tilde{f}$ corresponds to the median flux value in the out-of-eclipse portions of the light curve and $m_T = 7.93\,$mag is the \TESS{} magnitude of \VCen{}.

\subsection{New spectroscopy}
\label{sec:spectro}
In this work we utilize the data gathered by \citet{2015MNRAS.448.1937C}, who calculated radial velocities of \VCen{} from spectra taken with PUCHEROS and CORALIE spectrographs. However, in order to refine the parameters of the system, and better constrain the outer orbit, we made additional spectroscopic observations of \VCen{} with the CHIRON instrument \citep{2012SPIE.8446E..0BS,2013PASP..125.1336T}, attached to the 1.5-m telescope of the SMARTS consortium, located in the Cerro Tololo Inter-American Observatory (CTIO) in Chile. We monitored the target between May and September 2019, taking six spectra in the ``fiber'' mode, which provides high efficiency and spectral resolution of $R\simeq28000$. The exposure time was set to 600~s for the first four observations, and 867~s for the remaining two. The resulting signal-to-noise ratio $S/N$ varied between 110-170, with the exception of the fourth spectrum ($S/N\simeq25$). We aimed for the $S/N$ higher than in previous PUCHEROS and CORALIE observations in order to search for signatures of the third body, but we have found nothing conclusive (see Section~\ref{sec:bf}).

Data reduction and spectra extraction were performed on-site with the dedicated pipeline \citep{2013PASP..125.1336T}, and barycentric corrections to time and velocity were calculated with the {\it bcvcor} task of the {\it rvsao} package under {\it IRAF}. The RVs were calculated with our own implementation of the TODCOR method \citep{1994Ap&SS.212..349M}, for which we used two synthetic ATLAS9 template spectra: $T_{\rm eff}=6300$~K, $v_{\rm rot}=25$~k\ms{} for the primary, and $T_{\rm eff}=4700$~K, $v_{\rm rot}=20$~k\ms{} for the secondary. Measurement errors were estimated with a bootstrap procedure \citep{2012MNRAS.425.1245H}, sensitive to $S/N$ and rotational broadening of spectral lines.


\section{Analysis}
\label{sec:analysis}

\subsection{Light-curve modelling}
\label{sec:LC_analysis}

For the light-curve analysis of the \TESS{} and \Solaris{} data, we used the latest version (v34) of the \JKTEBOP{}\footnote{\url{https://www.astro.keele.ac.uk/~jkt/codes.html}} code \citep{2004MNRAS.351.1277S,2004MNRAS.355..986S}, which is based on the  Eclipsing Binary Orbit Program (\EBOP{}; \citealt{1981AJ.....86..102P}). We chose to fit the \TESS{} and \Solaris{} light curves using the same code as \citet{2015MNRAS.448.1937C} in order to compare and combine the results from different surveys. Although the current version of \JKTEBOP{} allows for fitting both the light curve and the radial velocity curves simultaneously, we did not use the radial-velocity data for the modeling. Indeed, the radial-velocity modulation of the eclipsing pair by a third body cannot be reproduced using the version 34 of \JKTEBOP{}. Moreover, the code is not designed to fit multiband light curves. In this model, the components are approximated as biaxial spheroids for the calculation of the reflection and ellipsoidal effects, and as spheres for the eclipse shapes. In addition, the \JKTEBOP{} code implements a Levenberg-Marquardt minimisation
method that allows to find the best-fit model parameters, including period $P$, time of primary minimum $T_0$, eccentricity $e$, argument of periastron $\omega$, inclination $i$, central surface brightness ratio $J$, sum of the fractional radii $r_1 + r_2$ (in units of semi-major axis $a$), and their ratio $k = r_2/r_1$. We also fitted the limb-darkening (LD) coefficients by adopting the logarithmic LD law proposed by \citet{1970AJ.....75..175K}. Initial values of the LD coefficients were taken from the tables of \citet{1993AJ....106.2096V}. We did not find evidence of a third light during the preliminary analysis of the \TESS{} and \Solaris{} data. As a result, the term $l_3/l_\mathrm{tot}$ was kept equal to zero when fitting the light curves.

\begin{figure}
	\includegraphics[trim = 0.0cm 1.3cm 1.0cm 1.3cm,clip,width=1.0\columnwidth,angle=0]{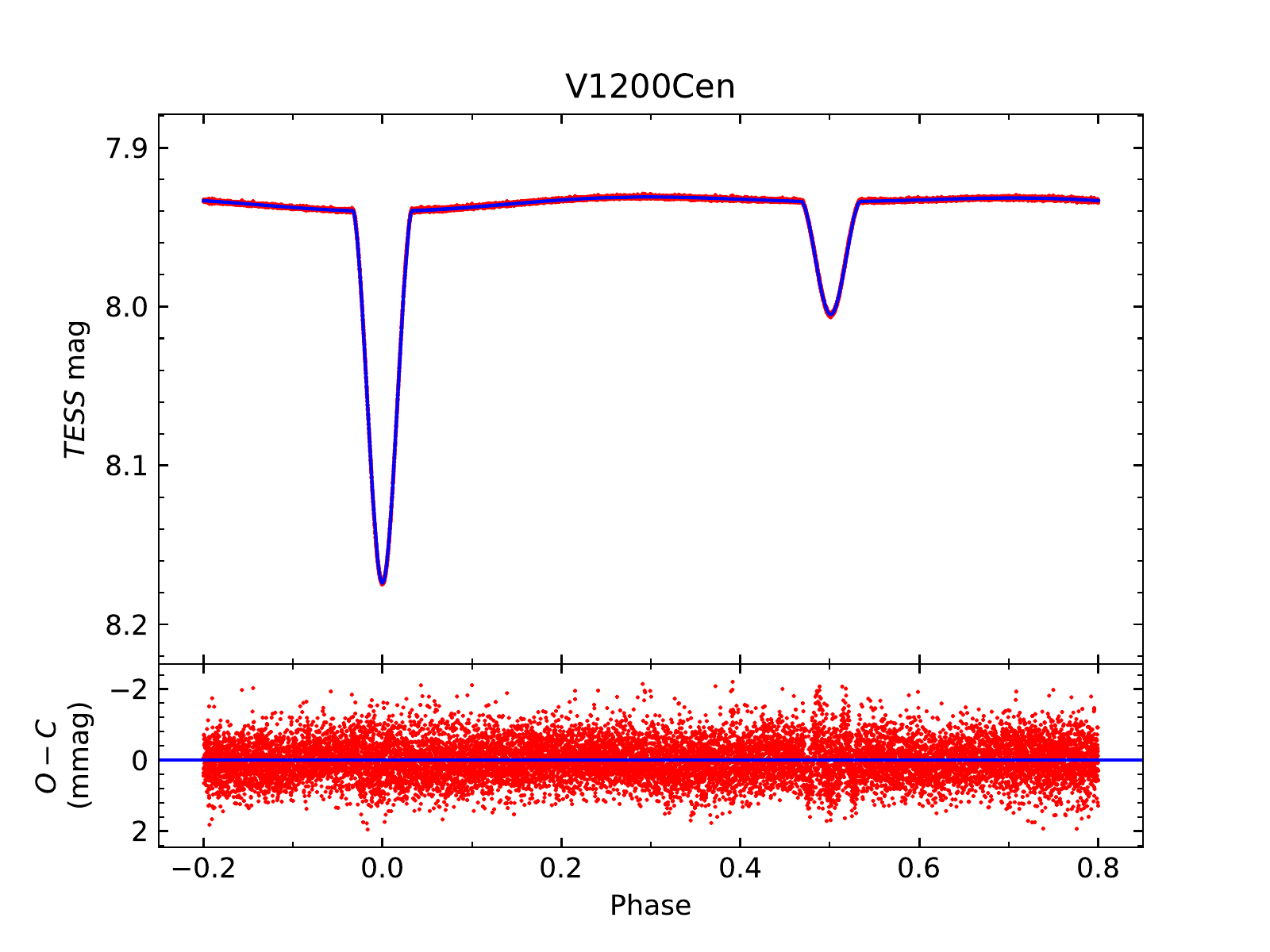}
	\includegraphics[trim = 0.0cm 0.0cm 1.0cm 1.3cm,clip,width=1.0\columnwidth,angle=0]{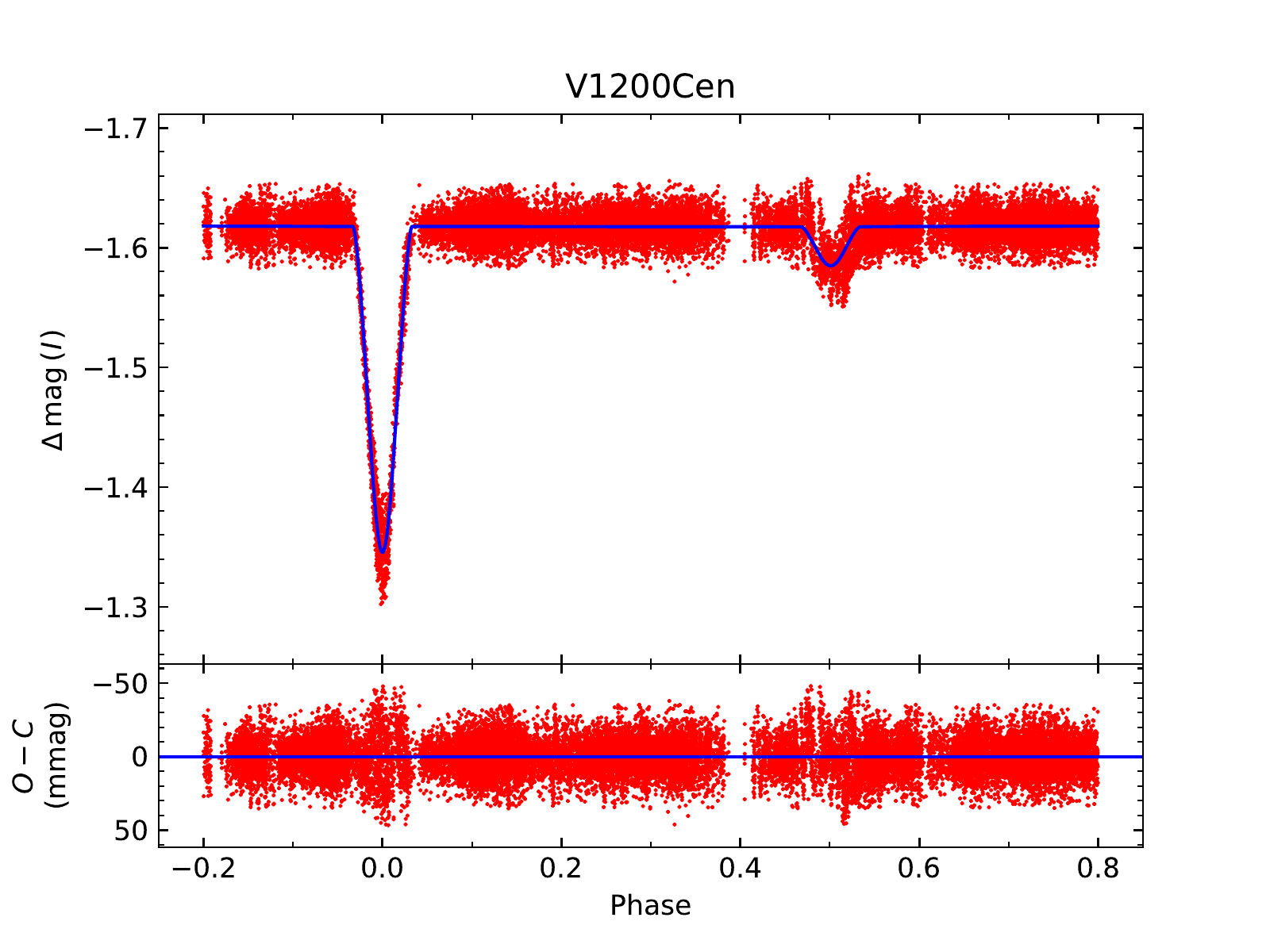}
    \caption{Phase-folded light curves of \VCen{} from \TESS{} (top) and from \Solaris{} in $I$ band (bottom). Red dots denote the observations while the blue line corresponds to the best-fit model obtained with \JKTEBOP{}. Fitting residuals are shown in lower panels. The $I$ band light curve was detrended before fitting in order to remove the long-term variability of the comparison star and the short-term atmospheric fluctuations. We used a detrending algorithm, namely W\={o}tan, which offers the possibility of masking the eclipses during the procedure. For the \TESS{} light curve, we proceeded in the way described in Section~\ref{sec:LC_analysis}. For graphical clarity, we removed the linear trend and the three sine waves from the \TESS{} data before plotting them.}
    \label{fig:LC_TESS_SLR_I}
\end{figure}

\begin{figure}
	\includegraphics[trim = 0.5cm 0.0cm 1.0cm 1.3cm,clip,width=1.0\columnwidth,angle=0]{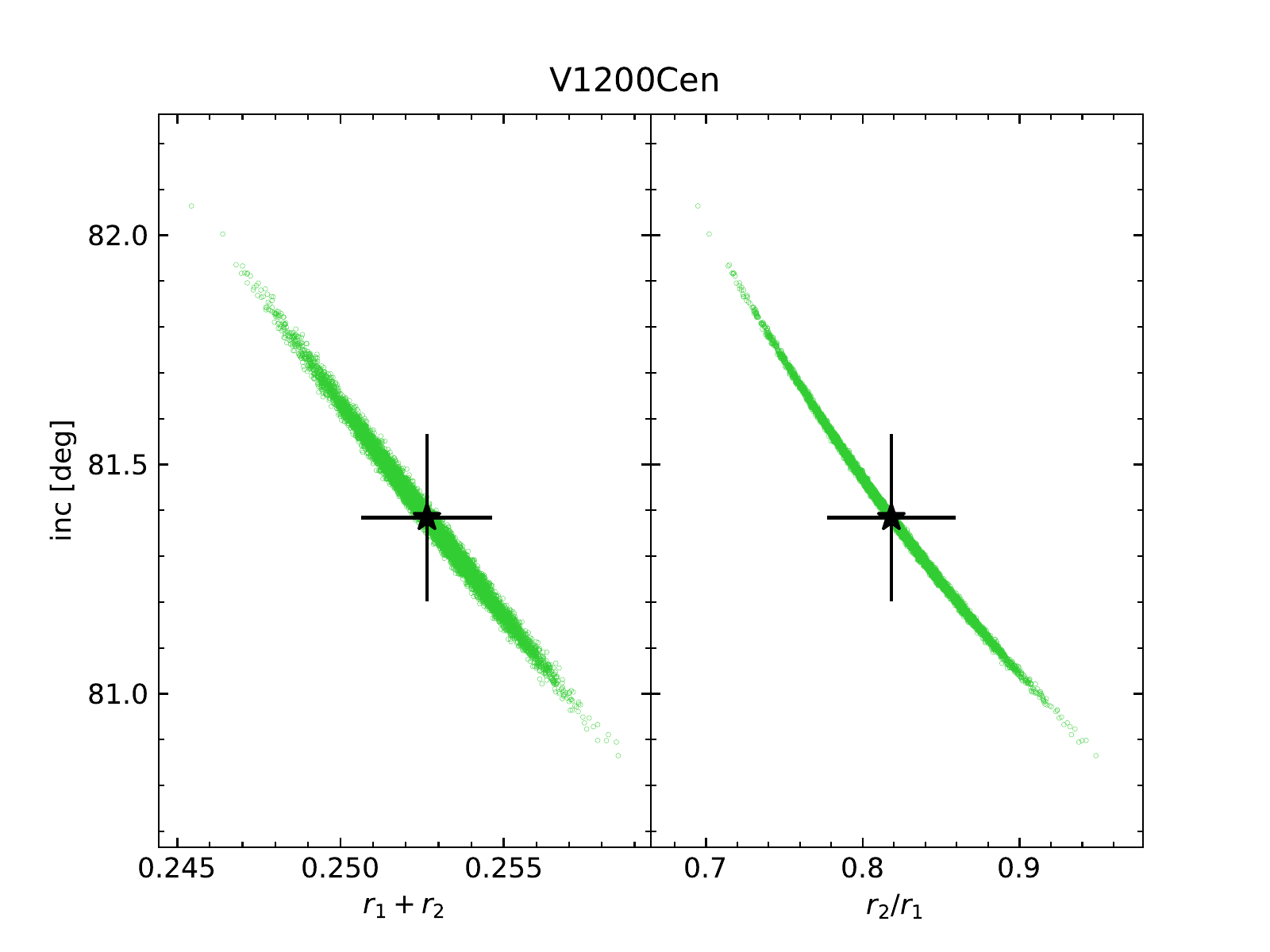}
    \caption{Results of the Monte Carlo analysis performed with \JKTEBOP{} on the \TESS{} data. Both plots present the distribution of the best-fit models in the $i$ versus $r_1+r_2$ (left) and $i$ versus $k=r_2/r_1$ (right) planes. Black stars with error bars indicate the mean values of the different parameters with their corresponding 1$\sigma$ uncertainties.}
    \label{fig:TESS-MC}
\end{figure}

Firstly, using \JKTEBOP{}, we performed the LC modelling of the high-precision \TESS{} photometric data. In addition to the model prescriptions detailed above, we also fitted the linear trend observed in the PDCSAP light curve as well as three sinusoidal curves with respective periods of ${\sim}P$, ${\sim}P/2$ and ${\sim}P/3$. For both the SAP and PDCSAP light curves, we identified the frequencies corresponding to ${\sim}P$, ${\sim}P/2$ and ${\sim}P/3$ in the Lomb-Scargle periodogram \citep{1976Ap&SS..39..447L,1982ApJ...263..835S}. We performed an initial fit of the light curves without sinusoidal modulation. Three sine waves with periods derived from the periodogram were then added, one by one, to the light curves. This allows to take the contribution of stellar activity, such as spots, into account. In this way, periodic variations with a semi-amplitude higher than about 0.7$\,$mmag were removed from the \TESS{} light curve, which is shown in Fig.~\ref{fig:LC_TESS_SLR_I}. For each model, we checked that the derived parameters are still consistent within their error bars. In contrast to the previous analysis of \citet{2015MNRAS.448.1937C}, the orbital eccentricity was found to be slightly different from zero. Indeed, a value of $e=0.01$ is required to properly fit the high-precision \TESS{} light curve. We then kept the eccentricity fixed to this value when modelling the \Solaris{} light curves. Furthermore, using \TESS{} photometry, we reduced by a factor of $\sim$7 the uncertainty on the orbital inclination such as derived by \citet{2015MNRAS.448.1937C}. This can be seen in Fig.~\ref{fig:TESS-MC} where we plotted the distribution of the inclination as a function of $r_1+r_2$ and $k$ (for a comparison, see Fig.~3 of \citealt{2015MNRAS.448.1937C}). Again, we adopted the value derived from \TESS{} photometry, $i = 81.38^\circ \pm 0.18^\circ$, as a fixed parameter in the model fitting of the \Solaris{} light curves. In Fig.~\ref{fig:LC_TESS_SLR_I}, we presented the observed \TESS{} and \Solaris{} light curves associated with their best-fit models derived from our analysis. The root mean squares (rms) of the residuals are $0.6\,$mmag and $12.0\,$mmag, respectively.

In order to derive reliable uncertainties, we performed Monte Carlo simulations on the \TESS{} and \Solaris{} light curves, with $10\,000$ runs each, as implemented in \JKTEBOP{} \citep{2005MNRAS.363..529S}. For the \Solaris{} light curves, the model parameters that were held fixed during the fitting procedure, \ie $e$ and $i$, were then perturbed in the Monte Carlo error analysis. Thus, the correlation between the fitted parameters can be assessed (see Fig.~\ref{fig:TESS-MC}). In Table~\ref{tab:MC_simu}, we summarised the results derived in this work, as well as those obtained by \citet{2015MNRAS.448.1937C} from the All-Sky Automated Survey (ASAS; \citealt{2002AcA....52..397P,2006MNRAS.368.1311P}) and Wide Angle Search for Planets (SuperWASP; \citealt{2006PASP..118.1407P}) surveys. We obtained very consistent results between the four surveys, based on various approaches. The main difference concerns the fractional radius of the secondary star, $r_2$, for which our estimate is about $4\%$ higher than the value adopted by \citet{2015MNRAS.448.1937C}. This is consistent with the correlations shown in Fig.~\ref{fig:TESS-MC}, where a lower inclination implies larger values of $r_1+r_2$ and $k$, and thus a larger fractional secondary radius. Thanks to the high precision of the \TESS{} photometry, we also significantly improved the precision on the fractional radii (${\sim}1\%$). From the fit of the SAP light curve, adopting the same approach as for the PDCSAP light curve, we obtained consistent results within their error bars, i.e.\ $r_1 = 0.137\,9\pm0.002\,9$, $r_2 = 0.116\,1\pm0.005\,6$ and $i = 81.27^\circ \pm 0.25^\circ$. Thus, the use of either the SAP or PDCSAP light curves does not change our conclusions. The implications of these results will be discussed in detail in Section~\ref{sec:discussion}.

\begin{table*}
\centering
\begin{minipage}{140mm}
	\caption{Parameters obtained from our analysis of the \TESS{} and \Solaris{} light curves and from the previous analysis of the ASAS and SuperWASP (SW) light curves by \citet{2015MNRAS.448.1937C}. The \Solaris{} values correspond to the $I$ band light curve shown in Fig.~\ref{fig:LC_TESS_SLR_I}. The values without error bars were held fixed during the fitting procedure. The adopted values are weighted means.}
	\label{tab:MC_simu}
	{
    \renewcommand{\arraystretch}{1.3}
	\begin{tabular}{@{}lccccc@{}} 
		\hline
		Parameter & \TESS{} value & \Solaris{} value & ASAS value & SW value & Adopted value \\
		\hline
		$P$ [d] & 2.482\,732\,7(69) & 2.482\,961\,5(44) & 2.482\,877\,8(43) &  2.482\,875\,2(25) & 2.482\,881\,1(19) \\
		$T_0$ [JD$-2\,450\,000$] & 1\,883.864\,432(39) & 1\,883.860\,791(91) & 1\,883.878\,9(31) & 1\,883.882\,7(24) & 1\,883.863\,863(36) \\
		$T$ [JD$-2\,450\,000$] & 1\,883.807\,765(39) & 1\,883.804\,266(91) & 1\,883.878\,9(31) & 1\,883.882\,7(24) & 1\,883.807\,212(36)$^\textit{a}$ \\
		$e$ & 0.010\,108\,0(25) & 0.010\,108\,0 & 0 & 0 & 0.010\,108\,0(25)$^\textit{b}$ \\
		$i$ [$^\circ$] & 81.38$\,\pm\,$0.18 & 81.38 & 81.9$^{+2.8}_{-1.3}$ & 81.6$^{+1.6}_{-1.3}$ & 81.38$\,\pm\,$0.18$^\textit{b}$ \\
		$r_1$ & 0.138\,9$\,\pm\,$0.001\,9 & 0.139\,1$\,\pm\,$0.001\,8 & 0.137$^{+0.014}_{-0.015}$ & 0.138$^{+0.025}_{-0.034}$ & 0.139\,0$\,\pm\,$0.001\,3 \\
		$r_2$ & 0.113\,7$\,\pm\,$0.003\,8 & 0.114\,1$\,\pm\,$0.001\,4 & 0.107$^{+0.024}_{-0.039}$ & 0.110$^{+0.038}_{-0.026}$ & 0.114\,0$\,\pm\,$0.001\,3$^\textit{c}$ \\
		\hline
	\end{tabular}
    }
    \textit{Notes}. $^\textit{a}$ The term $T$ corresponds to the time of periastron passage, which is different to the time of primary minimum $T_0$ when $e \neq 0$. The adopted value of $T$ was thus computed as the weighted mean of the \TESS{} and \Solaris{} values only. The values were shifted by $n\bar{P}$, where $n$ is an integer and $\bar{P}$ corresponds to the mean period derived from the different surveys.\\
    $^\textit{b}$ For $e$ and $i$, we adopted the well-constrained values from \TESS{}.\\
    $^\textit{c}$ The adopted value of $r_2$ was computed as the weighted mean of the \TESS{} and \Solaris{} values (see text).
\end{minipage}
\end{table*}

\subsection{Radial velocities and spectroscopic orbit}

In this section, we present the methodology employed to process the radial-velocity (RV) measurements of \VCen{}. Our analysis is based on the previously published RVs from the PUCHEROS\footnote{For more
details see \citet{2012MNRAS.424.2770V}.} and CORALIE\footnote{For more details see \citet{2001Msngr.105....1Q}.} spectrographs \citep{2015MNRAS.448.1937C}, supplemented by our own measurements collected with CHIRON (see Section~\ref{sec:spectro}). All previous and new RV measurements are provided in Table~\ref{tab:RV_obs} in the Appendix.

Recently, \citet{2018A&A...617A...2M} derived the orbit of the \Kepler{} target HD~188753 by applying a Bayesian analysis to the astrometric and RV measurements of the system. We adapted this Bayesian approach in order to fit a double-Keplerian orbit using the available RV data for \VCen. Thus, we defined the likelihood of the RV data given the orbital parameters as: 
\begin{equation}
    \ln {\cal L}_{\rm RV} = -\frac{1}{2}\,\sum^{N_{\rm RV}}_{i=1} \, \left(\frac{V_i^{\rm mod}-V_i^{\rm obs}}{\sigma_{V,i}} \right)^2,
\end{equation}
where $N_{\rm RV}$ denotes the number of observed radial velocities $V^{\rm obs}$ and $\sigma_V$ refers to the associated uncertainties. Here, the term $V^{\rm mod}$ corresponds to the RV values computed using an adapted version of the observable model described in \citet{2018A&A...617A...2M}. This new version takes into account the third-body perturbation by considering the orbital parameters $\mathcal{P}_{\rm orb} = (K_{Aa},K_{Ab},P_A,T_A,e_A,\omega_A,K_{A},P_{AB},T_{AB},e_{AB},\omega_{AB},\gamma_{AB})$, where subscripts Aa and Ab refer respectively to the primary and secondary components and subscript AB refers to the relative orbit between the eclipsing pair A and the third body B. We use this notation from now on, which is more adequate to describe a possible quadruple star system. For the derivation of these orbital parameters, we employed a Markov chain Monte Carlo (MCMC) method using the Metropolis-Hastings algorithm (MH; \citealt{1953JChPh..21.1087M,hastings70}) as explained in \citet{2018A&A...617A...2M}. Briefly, the procedure consists of setting ten chains of 10 million points each with starting points taken randomly from appropriate distributions. The new
set of orbital parameters is here computed using a random walk as:
\begin{equation}
{\cal P}_{\rm orb}^{t'}={\cal P}_{\rm orb}^{t}+\alpha_{\rm rate} \, \Delta {\cal P}_{\rm orb},
\end{equation}
where $\Delta {\cal P}_{\rm orb}$ is given by a multinomial normal distribution with independent parameters and $\alpha_{\rm rate}$ is an adjustable parameter that is reduced by a factor of two until the rate of acceptance of the new set exceeds 25$\%$. We then derived the posterior probability of each parameter from the chains after rejecting the initial burn-in phase (\ie the first 10$\%$ of each chain). For all parameters, we computed the median and the credible intervals at 16$\%$ and 84$\%$, corresponding to a 1$\sigma$ interval for a normal distribution. An essential aspect when fitting data from different instruments is
the determination of a proper relative weighting. To this end, we performed a preliminary set of fits and calculated the rms of the residuals for stars Aa and Ab. We obtained rms values of 0.76 and 1.80\,km\,s$^{-1}$, respectively, which were adopted as weights for the final fit. We then checked that the rms values from our best solution are equal to the weights used during the fitting procedure.

\begin{figure*}
	\includegraphics[trim = 0.0cm 2.0cm 0.0cm 0.5cm,clip,width=0.85\columnwidth,angle=90]{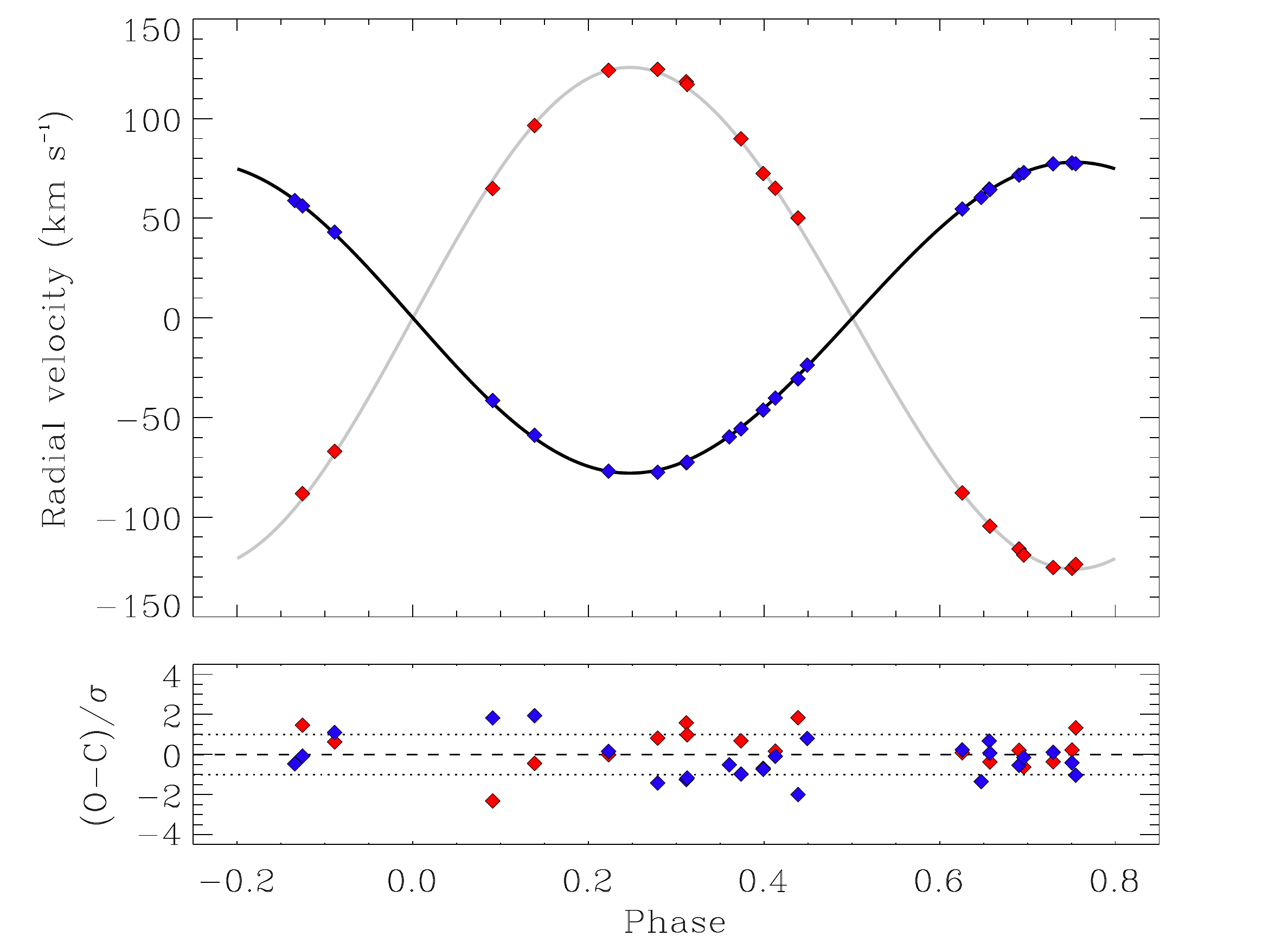}
	\includegraphics[trim = 0.0cm 2.0cm 0.0cm 0.5cm,clip,width=0.85\columnwidth,angle=90]{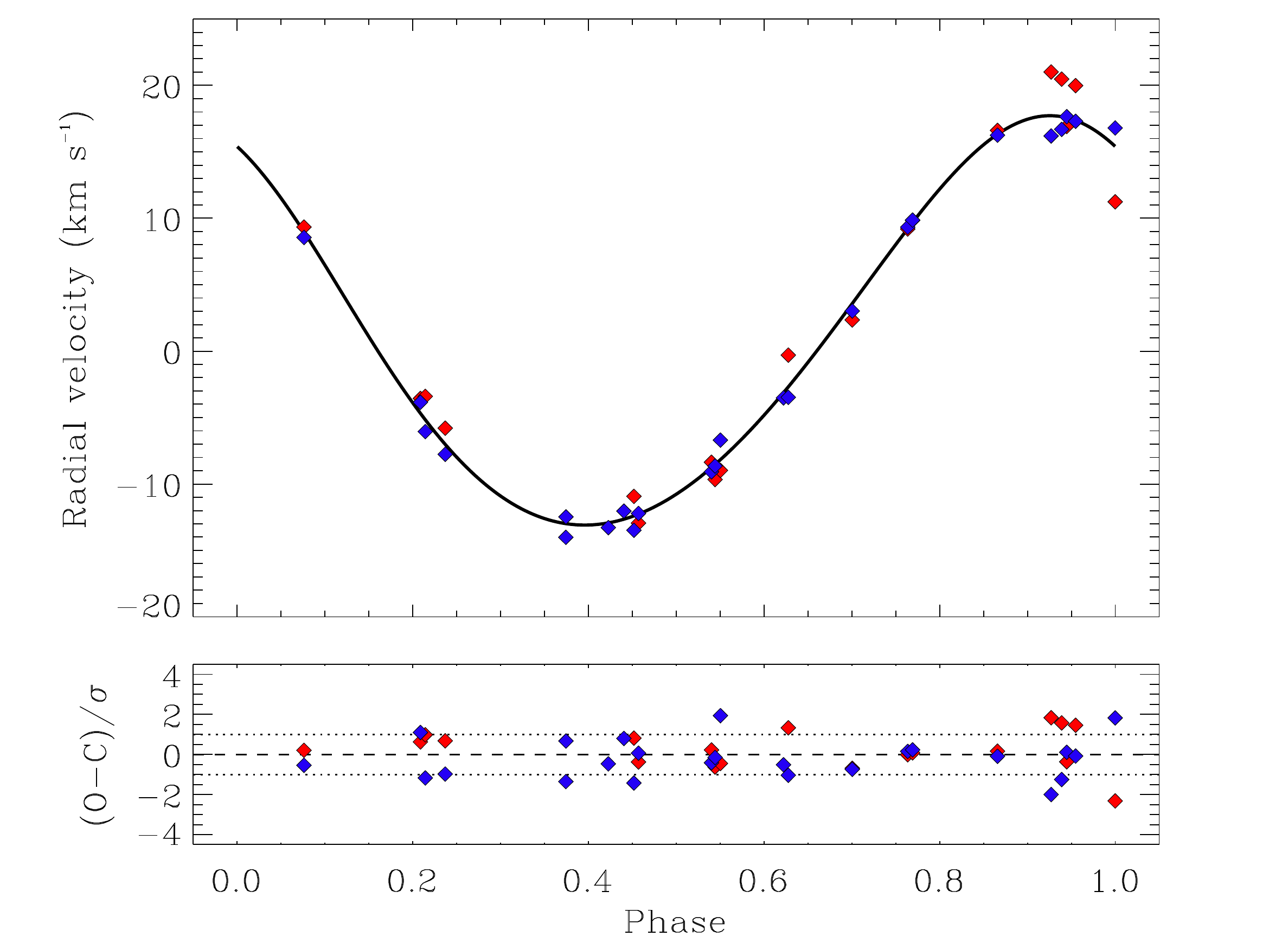}
    \caption{RV curves of \VCen{} described by a double-Keplerian orbital model using radial velocities of stars Aa (blue) and Ab (red). Upper-left panel: Best-fit solutions for stars Aa (black) and Ab (grey) after having removed the 180-day modulation induced by the third body. The curve is phase-folded at the orbital period $P_A\!\simeq 2.5\,$days, where phase 0 is set for the time of primary minimum $T_0$. Upper-right panel: Best-fit solution for the centre of mass of the eclipsing pair after having removed the orbital motion of stars Aa and Ab. The curve is phase-folded at the orbital period $P_{AB}\!\simeq 180\,$days, where phase 0 is set for the time of periastron passage $T_{AB}$. Lower panels: Residuals of the fitting procedure.}
    \label{fig:RV_phase}
\end{figure*}

Finally, the best-fit RV solution for the whole system is shown in Fig.~\ref{fig:RV_phase}. We also indicated the corresponding orbital parameters derived from our Bayesian analysis of the RV data in Table~\ref{tab:RV_param}. We fitted two additional terms to take into account the zero-point differences, \ie E/C$-$5/P and C/C$-$5/P. Here, we assumed that the shift is the same for the two stars. We then obtained E/C$-$5/${\rm P} = -0.23\pm0.67$\,km\,s$^{-1}$ and C/C$-$5/${\rm P} = -0.27\pm0.73$\,km\,s$^{-1}$. The new values of the orbital and physical parameters are, within the error bars, in agreement with those presented in Table~\ref{tab:RV_param}. Therefore, in the following, we adopted as reference the solution where E/C$-$5/P and C/C$-$5/P are equal to zero. Thanks to the new RV measurements from CHIRON, we obtained a more precise and robust solution than that of \citet{2015MNRAS.448.1937C}. Indeed, we found that the AB system has a 180-day orbital period and is almost circular with an eccentricity of 0.088. For comparison, \citet{2015MNRAS.448.1937C} derived an orbital period of 351.5~days and an eccentricity of 0.42 for the AB system. Such a difference is due to the fact that the authors did not have enough RV data points to uniformly cover the orbit of the AB system due to uneven sampling, while the time span of our CHIRON observations is comparable to $P_{AB}$, and sampling was relatively regular. The new orbit implies a more massive third body than in the previous study, namely $M_B = 0.871\,{\rm M}_\odot$ instead of 0.662$\,{\rm M}_\odot$ (minimum mass for the third body, corresponding to $i_{AB}\!= 90^\circ$), whose implications will be further discussed in Section~\ref{sec:discussion}.  

\begin{table}
\centering
\begin{minipage}{85mm}
	\caption{Orbital parameters and derived quantities for the best-fit model of the RV data.}
	\label{tab:RV_param}
	{
    \renewcommand{\arraystretch}{1.0}
	\begin{tabular}{@{}lccc@{}}
		\hline
		Parameter & Median & 84$\%$ interval & 16$\%$ interval \\
		\hline
		$K_{Aa}$ [km\,s$^{-1}$] & 78.02  & +0.28 & $-$0.28 \\
		$K_{Ab}$ [km\,s$^{-1}$] & 125.89 & +0.70 & $-$0.71 \\
		$P_A$ [d] & 2.482\,881\,1$^\textit{a}$ & \multicolumn{2}{c}{$\pm$0.000\,001\,9} \\
		$T_A$ [JD$-2\,450\,000$] & 1\,883.807\,212$^\textit{a}$ & \multicolumn{2}{c}{$\pm$0.000\,036} \\
		$e_A$ & 0.010\,108\,0$^\textit{a}$ & \multicolumn{2}{c}{$\pm$0.000\,002\,5} \\
		$\omega_A$ [$^\circ$] & 81.616$^\textit{a}$ & \multicolumn{2}{c}{$\pm$0.096} \\
		$K_A$ [km\,s$^{-1}$] & 15.41 & +0.27 & $-$0.27 \\
		$P_{AB}$ [d] & 180.374 & +0.093 & $-$0.094 \\
		$T_{AB}$ [JD$-2\,450\,000$] & 5\,818.7 & +7.1 & $-$6.8 \\
		$e_{AB}$ & 0.088 & +0.018 & $-$0.019 \\
		$\omega_{AB}$ [$^\circ$] & 32 & +13 & $-$13 \\
		$\gamma_{AB}$ [km\,s$^{-1}$] & 1.16 & +0.22 & $-$0.22 \\
		\hline
		$a_{Aab}\sin i_A$ [R$_\odot$]$^\textit{b}$ & 10.006 & +0.037 & $-$0.037 \\
		$q$ & 0.619\,7 & +0.004\,2 & $-$0.004\,2 \\
		$M_{Aa}\sin^3 i_A$ [M$_\odot$] & 1.346 & +0.017 & $-$0.017 \\
		$M_{Ab}\sin^3 i_A$ [M$_\odot$] & 0.834\,4 & +0.007\,8 & $-$0.007\,8 \\
		$a_{A}\sin i_{AB}$ [au]$^\textit{b}$ & 0.254\,4 & +0.004\,4 & $-$0.004\,5 \\
		$f(M_B)$ [M$_\odot$] & 0.067\,5 & +0.003\,6 & $-$0.003\,5 \\
		$M_B \, (i_{AB}\!=90^\circ$) [M$_\odot$] & 0.871 & +0.020 & $-$0.020 \\
		\hline
	\end{tabular}
	}
    \textit{Notes}. $^\textit{a}$ During the fitting procedure, these four parameters are drawn from a normal distribution centered on the values derived using \JKTEBOP{}, with a dispersion equal to their 1$\sigma$ uncertainties.\\
    $^\textit{b}$ For the eclipsing pair, we differentiate the semi-major axis $a_{Aab}\!=a_{Aa}\!+a_{Ab}$ of the relative orbit from the semi-major axis $a_A\!=a_{AB}\!-a_{B}$ of the barycentric orbit. Their respective inclinations are $i_A\!=81.38^\circ$ (see Table~\ref{tab:MC_simu}) and $i_{AB}\!=90^\circ$ (assumed).
\end{minipage}
\end{table}

\subsection{Eclipse timing variations}
\label{sec:etv}

It is well known that EBs can exhibit period changes as a result of the gravitational attraction of a third body through the light-travel-time effect (LTTE; \citealt{1990BAICz..41..231M}), also known as the R{\o}mer delay. Furthermore, since the orbital motion is not purely Keplerian in a multi-body case, the EB undergoes a number of dynamical perturbations \citep{2013ApJ...768...33R,2015MNRAS.448..946B,2016MNRAS.455.4136B}, of which the strongest are those with time scales of $P_{AB}$. Both effects result in eclipse timing variations (ETVs), with amplitudes dependent on the parameters of the outer body. In the following, we will take advantage of the high-precision photometric data collected by \TESS{} to perform the ETV analysis of \VCen{}.

\subsubsection{$O-C$ eclipse times}
\label{sec:ecl_times}

\begin{figure*}
	\includegraphics[trim = 3.0cm 2.0cm 3.0cm 2.0cm,clip,width=1.0\columnwidth,angle=0]{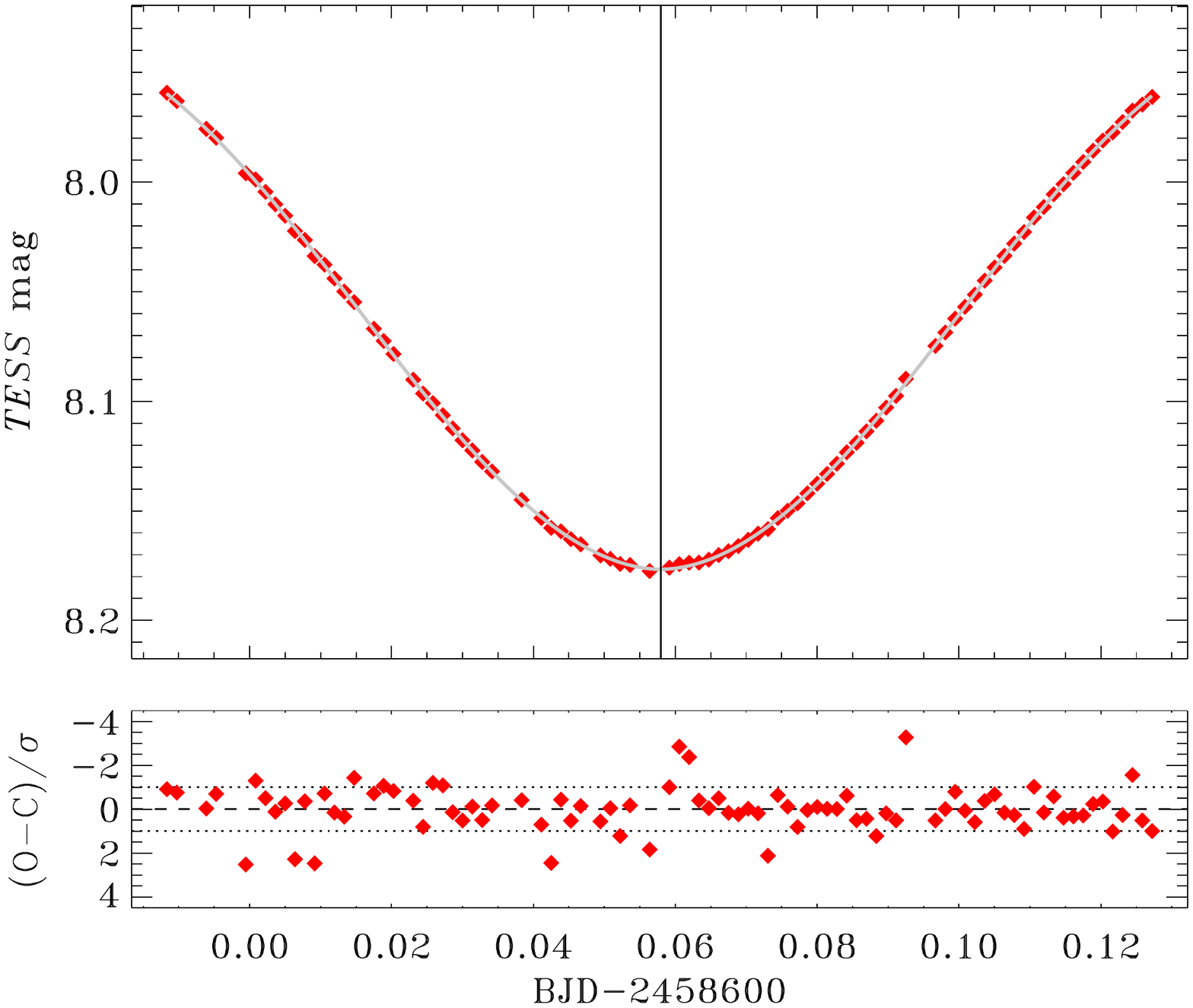}
	\includegraphics[trim = 3.0cm 2.0cm 3.0cm 2.0cm,clip,width=1.0\columnwidth,angle=0]{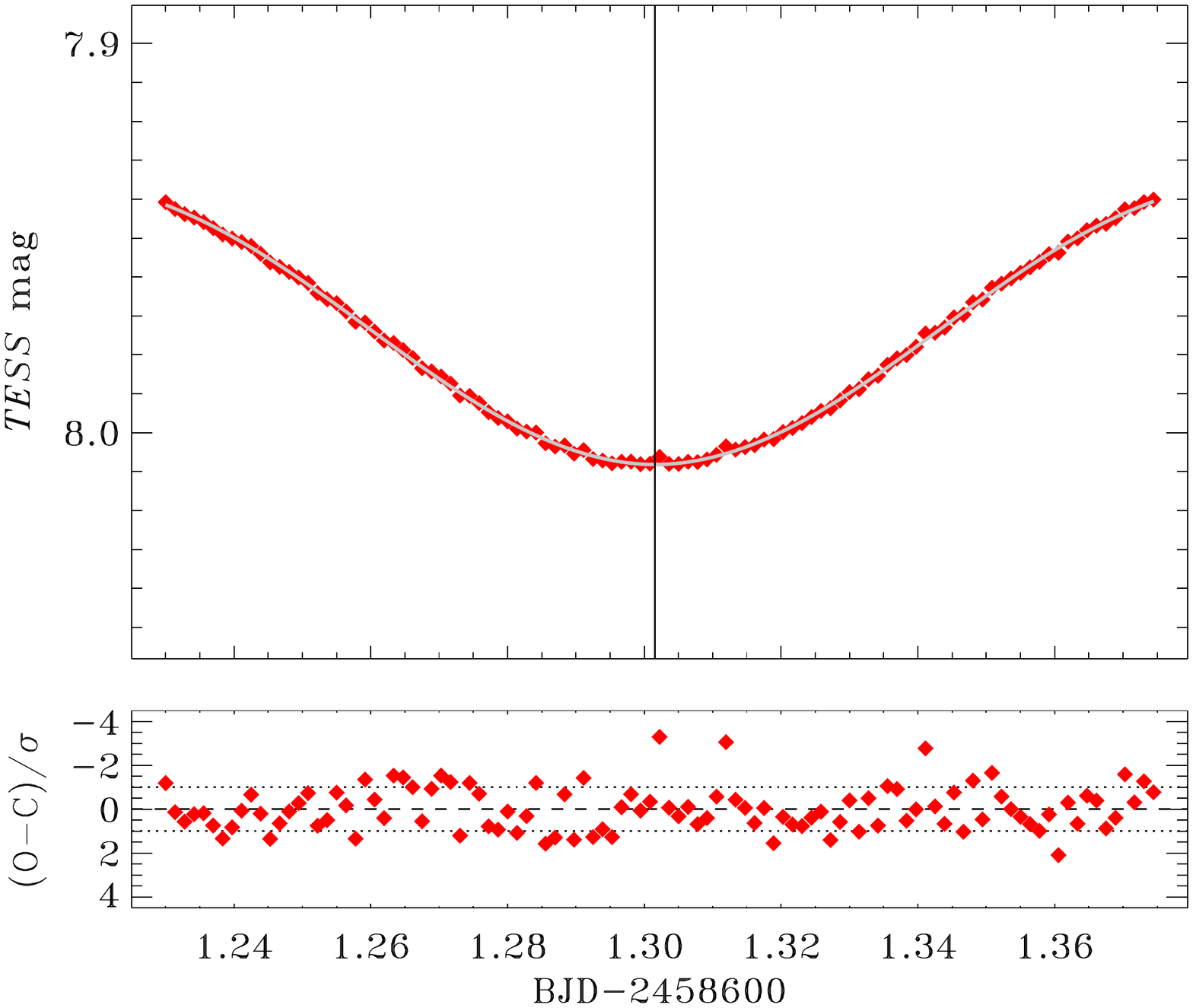}
    \caption{
	\TESS{} photometry of the primary (left) and secondary (right) eclipses of \VCen{}. Red diamonds denote the observations while the grey line corresponds to the best-fit model determined for each eclipse using the procedure described in Section~\ref{sec:ecl_times}. The vertical line indicates the time of minimum light associated with the best-fit model (see Table~\ref{tab:ecl_times}). Fitting residuals are shown in lower panels.
	}
    \label{fig:ecl_times}
\end{figure*}

As a first step of the ETV analysis, we determined the times of minima of the primary and secondary eclipses from the \TESS{} PDCSAP light curve shown in Fig.~\ref{fig:TESS-PDCSAP}. To this end, we adopted the formalism of \citet{2015A&A...584A...8M} that describes the morphology of the eclipse profile. The corresponding model is defined as:  
\begin{equation}
    f(t_i,\theta) = \alpha_0 + \alpha_1 \, \psi(t_i,t_0,d,\Gamma),
    \label{eq:ecl_model_1}
\end{equation}
where $\alpha_0$ is the magnitude zero-point shift and $\alpha_1$ is a multiplicative constant of the eclipse profile function, which is written as:
\begin{equation}
    \psi(t_i,t_0,d,\Gamma) = 1-\bigg\{ 1-\exp \bigg[ 1-\cosh \bigg( \frac{t_i-t_0}{d} \bigg) \bigg] \bigg\}^\Gamma.
    \label{eq:ecl_model_2}
\end{equation}
Here, $t_0$, $d$ and $\Gamma$ are the time of minimum, the eclipse width and the kurtosis, respectively. Each eclipse is thus described by the following parameters $\theta = (\alpha_0,\alpha_1,t_0,d,\Gamma)$. We then performed an MCMC fit of each eclipse using the model from equations~(\ref{eq:ecl_model_1}) and~(\ref{eq:ecl_model_2}). The best-fit solutions obtained for two consecutive eclipses (primary and secondary) are shown in Fig.~\ref{fig:ecl_times}. The rms of the residuals for the different fits is between 0.5 and 0.7$\,$mmag, consistent with the value from the global fit (see Section~\ref{sec:LC_analysis}). Table~\ref{tab:ecl_times} summarises the times of minima derived from our fitting procedure. The associated uncertainties are of the order of 2 and 6$\,$s for the primary and secondary eclipses, respectively.

The second step of the ETV analysis consists on computing the $O-C$ residuals between the observed and calculated times of minima, i.e.:
\begin{equation}
    \Delta = T_{\rm o}(E) - T_{\rm c}(E) = T_{\rm o}(E) - T_0 - P E,
    \label{eq:OC_times}
\end{equation}
where $T_{\rm o}(E)$ and $T_{\rm c}(E)$ refer to the observed and calculated times of minima at epoch $E$, respectively. The values of $T_0$ and $P$ are taken from Table~\ref{tab:MC_simu}. For secondary eclipses, the term $-\,(T_2-T_1)$ has to be added to the right side of equation~(\ref{eq:OC_times}). It corresponds to the time interval between the primary and secondary eclipses (\citealt{1959cbs..book.....K,2001icbs.book.....H}):
\begin{equation}
    \frac{2 \pi \, (T_2-T_1)}{P} = \pi + 2 \tan^{-1} \frac{e \cos \omega}{(1-e^2)^{1/2}} + \frac{2 e \cos \omega \, (1-e^2)^{1/2}}{(1-e^2 \sin^2 \omega)},
\end{equation}
where $e$ and $\omega$ are the eccentricity and the argument of periastron of the eclipsing pair, respectively (see Tables~\ref{tab:MC_simu} and~\ref{tab:RV_param}). \citet{2013ApJ...774...81T} showed that the primary and secondary times of minimum light can be affected differently by starspots, making their respective ETVs to be out of phase with each other. As suggested by these authors, we adopted the averaged $O-C$ values between each pair of consecutive primary and secondary minima in order to better reproduce the contribution of the third body.

\begin{table}
\centering
\begin{minipage}{45mm}
	\caption{
	Times of minima of the primary and secondary eclipses from the \TESS{} light curve of \VCen{}.
	}
	\label{tab:ecl_times}
	{
    \renewcommand{\arraystretch}{1.0}
	\begin{tabular}{@{}lcc@{}}
		\hline
		Time & Cycle & 1$\sigma$ error \\ \relax
		[BJD$-2\,450\,000$] & no. & [d] \\
		\hline
		8\,600.057\,939 & 0.0 & 0.000\,026 \\
		8\,601.301\,580 & 0.5 & 0.000\,082 \\
        8\,602.540\,706 & 1.0 & 0.000\,023 \\
        8\,603.784\,187 & 1.5 & 0.000\,074 \\
        8\,605.023\,414 & 2.0 & 0.000\,026 \\
        8\,606.266\,815 & 2.5 & 0.000\,082 \\
        8\,607.506\,084 & 3.0 & 0.000\,019 \\
        8\,608.749\,451 & 3.5 & 0.000\,062 \\
        8\,614.954\,306 & 6.0 & 0.000\,028 \\
        8\,616.197\,696 & 6.5 & 0.000\,072 \\
        8\,617.437\,034 & 7.0 & 0.000\,021 \\
        8\,618.680\,392 & 7.5 & 0.000\,074 \\
        8\,619.919\,768 & 8.0 & 0.000\,021 \\
        8\,621.163\,152 & 8.5 & 0.000\,069 \\
        8\,622.402\,544 & 9.0 & 0.000\,029 \\
		8\,623.645\,822 & 9.5 & 0.000\,067 \\
		\hline
	\end{tabular}
	}
	\textit{Notes}. Half-integer cycle numbers refer to secondary eclipses. There are no eclipses observed for cycle nos.\ 4.0 to 5.5 (see text). 
\end{minipage}
\end{table}

\subsubsection{LTTE ETV solution}
\label{sec:ltte_etv}

According to \citet{1952ApJ...116..211I}, the semi-amplitude of the LTTE ETV is given by:
\begin{equation}
    A_{\rm LTTE} = \frac{a_{A}\sin i_{AB}}{c},
\end{equation} 
where $c$ is the speed of light. In the case of \VCen, we have $a_{A}\sin i_{AB} = 0.254\,4\,$au (see Table~\ref{tab:RV_param}), implying that ETVs can be seen with a semi-amplitude of about 127$\,$s. It is worth mentioning that the RVs of the centre-of-mass of the inner EB are simple time derivatives of the R{\o}mer delay in ETVs, only multiplied by the speed of light $c$ \citep{2016MNRAS.461.2896H}.

\begin{figure}
	\includegraphics[trim = 2.5cm 2.0cm 3.0cm 2.0cm,clip,width=1.0\columnwidth,angle=0]{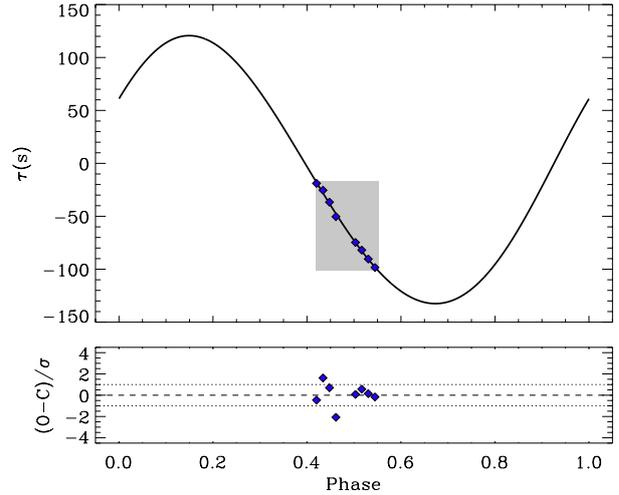}
    \caption{
	ETV curve of \VCen{}. Blue diamonds denote the time residuals, derived as explained in Section~\ref{sec:ecl_times}, while the black line corresponds to the best-fit solution of the ETV model described by equation~(\ref{eq:etv_model}), considering only the LTTE. The corrections were applied to the measurements, for clarity purpose. The grey area indicates the phase range covered by \TESS{} observations. Fitting residuals are shown in the lower panel.
	}
    \label{fig:TESS-ETV}
\end{figure}

As it will be shown in the next section, the LTTE can no longer be considered as the only source of the ETV signal observed for \VCen{}. Nevertheless, for completeness, we chose to present the LTTE model without additional effects. We then modelled the ETV in the following mathematical form \citep{2015MNRAS.448..946B,2016MNRAS.455.4136B}:
\begin{equation}
    \Delta = c_0 + c_1 E - \frac{a_{A}\sin i_2}{c} \, \frac{\big(1-e_2^2\big)\sin(\nu_2+\omega_2)}{1+e_2\cos \nu_2},
    \label{eq:etv_model}
\end{equation}
where $c_0$ and $c_1$ are factors that correct the respective values of $T_0$ and $P$ for the ETV effect. Here, the subscript `2' refers to the outer orbit, namely the AB system. We then searched for the values of $c_0$ and $c_1$ that best fit the averaged $O-C$ residuals, as defined above, using an MCMC procedure. During the fitting process, the outer orbital parameters were fixed at their values determined from the RV measurements and listed in Table~\ref{tab:RV_param}. The model ETV curve is shown in Fig.~\ref{fig:TESS-ETV} with the observed $O-C$ residuals overplotted. The rms of the fit is 1.5$\,$s. Finally, the correction factors were found to be $c_0 = 69.9\pm1.4\,$s and $c_1 = -4.5\pm0.3\,$s. 

\subsubsection{Combined dynamical and LTTE ETV solution}

The ETV semi-amplitude of the second effect, very often called ``physical'' or ``dynamical'' in the literature, for circular EB orbits is given by \citep{2016MNRAS.455.4136B}:
\begin{equation}
    A_{\rm phys} = \frac{3}{4\pi} \frac{M_B}{M_A+M_B} \frac{P_1^2}{P_2} \big(1-e_2^2\big)^{-3/2},
\end{equation}
where the subscripts `1' and `2' refer to the inner and outer orbits, respectively. We have the lower mass limit of $M_B$ (0.871$\,$M$_\odot$). Thus, we can estimate that $A_{\rm phys}$ is not smaller than 197$\,$s, making it also a non-negligible effect. Furthermore, if we consider that the B sub-system is composed of two stars with masses lower than $M_{Ab} = 0.863\,3\,$M$_\odot$ (see Section~\ref{sec:bf}), then we can put an upper limit to its total mass of 1.727$\,$M$_\odot$, corresponding to a maximum amplitude of 309$\,$s.

\begin{figure*}
	\includegraphics[trim = 2.5cm 2.0cm 3.0cm 2.0cm,clip,width=1.0\columnwidth,angle=0]{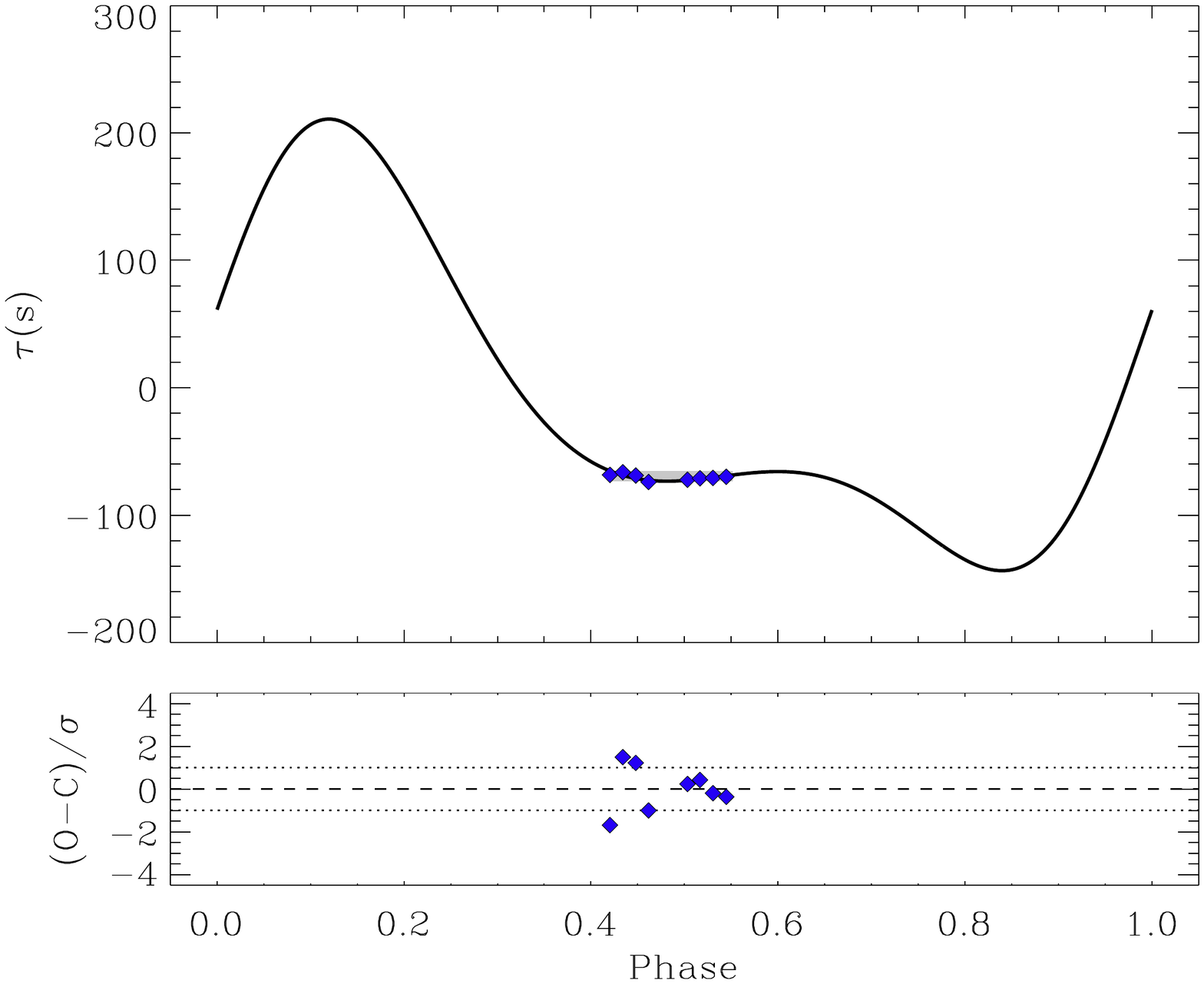}
	\includegraphics[trim = 2.5cm 2.0cm 3.0cm 2.0cm,clip,width=1.0\columnwidth,angle=0]{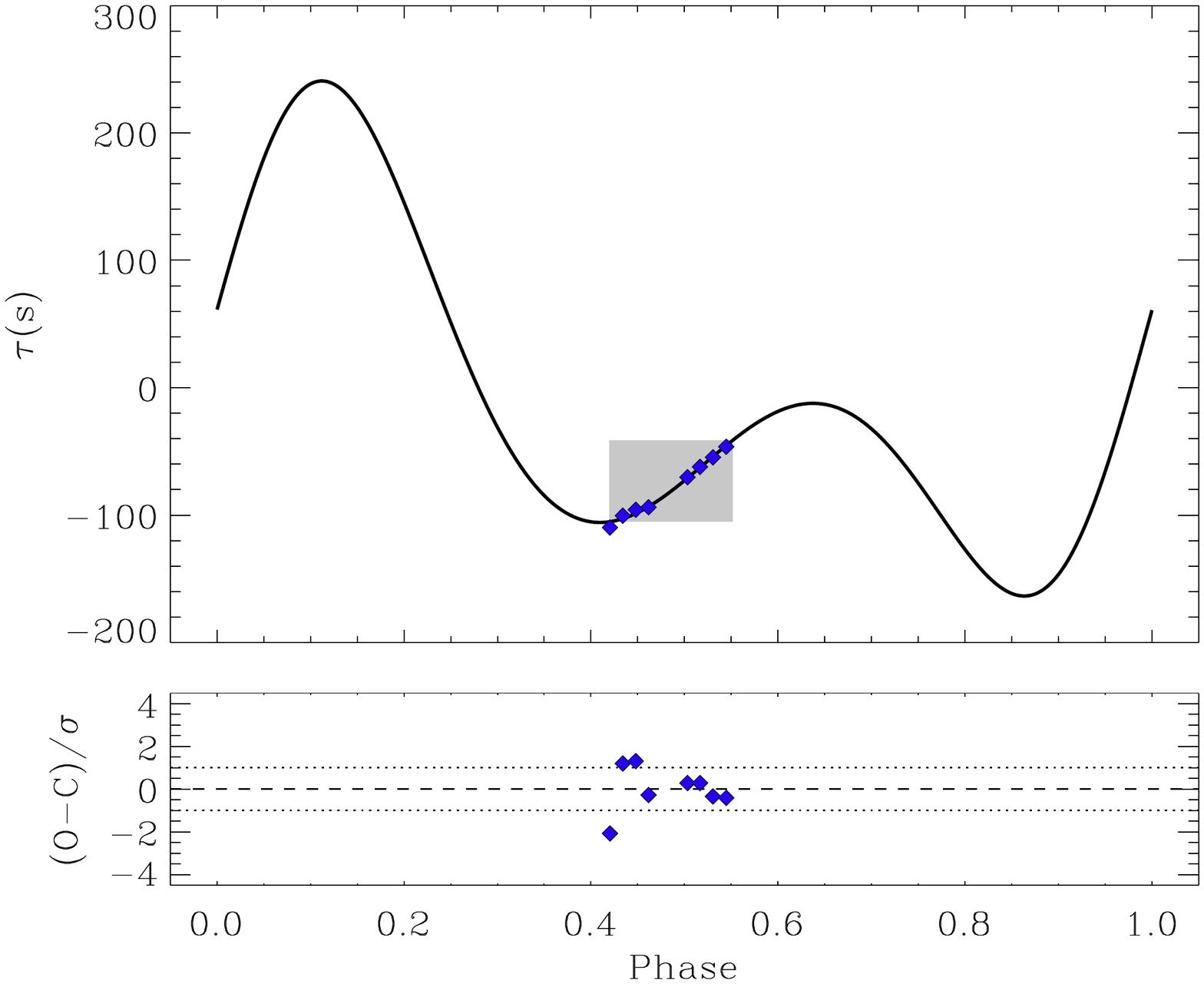}
    \caption{
    Same as Fig.~\ref{fig:TESS-ETV}, but taking into account both the LTTE and dynamical perturbation effects for two different values of the mutual inclination, namely $i_{\rm m}\!=45.1^\circ$ (left) and $i_{\rm m}\!=117.7^\circ$ (right). More details are given in the text.
    }
    \label{fig:TESS-ETV-dyn}
\end{figure*}

The exact formula for the physical delay in time includes the mutual inclination angle between outer and inner orbits \citep{2013ApJ...768...33R,2015MNRAS.448..946B,2016MNRAS.455.4136B}, which is unknown in our case. Also, on the contrary to the LTTE, this effect can not be directly derived from the RVs. However, the time coverage of the \TESS{} observations is long enough to see the ETVs manifest in the light curve, as shown in the right-hand panel of Fig.~\ref{fig:TESS-ETV-dyn}. The ETV contribution of the dynamical perturbations takes the following form \citep{2016MNRAS.455.4136B}:
\begin{equation}
    \Delta_{\rm phys} = A_{\rm phys} \bigg[ \bigg( \frac{2}{3}-\sin^2 i_{\rm m} \bigg) {\cal M} + \frac{1}{2} \sin^2 i_{\rm m} {\cal S} \bigg],
    \label{eq:dyn_model_1}
\end{equation}
where
\begin{equation}
    {\cal M} = 3 e_2 \sin \nu_2 - \frac{3}{4} e_2^2 \sin 2\nu_2 + \frac{1}{3} e_2^3 \sin 3\nu_2 + {\cal O}(e_2^4),
    \label{eq:dyn_model_2}
\end{equation}
and
\begin{equation}
    {\cal S} = \sin(2\nu_2+2g_2) +e_2 \bigg[ \sin(\nu_2+2g_2) + \frac{1}{3} \sin(3\nu_2+2g_2) \bigg].
    \label{eq:dyn_model_3}
\end{equation}
Equations~(\ref{eq:dyn_model_1}) to~(\ref{eq:dyn_model_3}) are defined for a circular inner orbit. This approximation appears reasonable here since the eccentricity is small ($e_1=0.01$). In addition, we only need to have a rough estimate of the correction factor to apply to the inner period and thus to the masses of the eclipsing components. By adopting the value of $M_B = 1.727\,$M$_\odot$, we can determine the inclination of the outer orbit from the mass function, which we found to be $i_{AB}\!=36.3^\circ$. The mutual inclination, $i_{\rm m}$, between the two orbital planes is given by \citep{1973bmss.book.....B,1981ApJ...246..879F}:
\begin{equation}
\cos i_{\rm m} = \cos i_1 \cos i_2+ \sin i_1 \sin i_2 \cos(\Omega_1-\Omega_2),\label{eq:mutual_inc}
\end{equation}
where $\Omega_1$ and $\Omega_2$ are the position angles of the line of nodes of the inner and outer orbits, respectively. In the case of \VCen{}, these two angles are unknown and thus the mutual inclination $i_{\rm m}$ cannot be determined. However, it is easy to show from equation~(\ref{eq:mutual_inc}) that $i_A - i_{AB} \leq i_{\rm m} \leq i_A + i_{AB}$. From the derived values of $i_A$ and $i_{AB}$, we then obtain a mutual inclination of $45.1^\circ \leq i_{\rm m} \leq 117.7^\circ$. The mutual inclination was set in equation~(\ref{eq:dyn_model_1}) to these limit values during the fit. The only unknown parameter is the dynamical argument of periastron ($g_2$), which was arbitrarily fixed at zero. We checked that varying this parameter does not change the order of magnitude of the corrections.

In order to take the dynamical effect into account, the dynamical perturbation term defined in equation~(\ref{eq:dyn_model_1}) was added to equation~(\ref{eq:etv_model}). For each of the two values of $i_{\rm m}$, we searched for the values of $c_0$ and $c_1$ that best fit the $O-C$ eclipse times, as detailed in Section~\ref{sec:ltte_etv}. The corresponding ETV curves are shown in Fig.~\ref{fig:TESS-ETV-dyn}. When both effects are simultaneously considered, we obtain $c_0 = 119.8\pm1.4\,$s and $c_1 = -13.1\pm0.3\,$s for $i_{\rm m} = 45.1^\circ$, and $c_0 = 161.4\pm1.7\,$s and $c_1 = -20.3\pm0.3\,$s for $i_{\rm m} = 117.7^\circ$. The rms are 1.7$\,$s and 2.2$\,$s, respectively. It results that, in the less favourable case, the period value adopted in Table~\ref{tab:MC_simu} for the eclipsing pair has to be shifted by $-20.3\,$s to account for ETVs. Such corrections represent a very small fraction of the period of the eclipsing pair. Thus, the systematic error on the masses $M_{Aa}$ and $M_{Ab}$ due to an incorrect estimation of the period should be of $\Delta P/P = 0.009\%$, which is largely below the precision claimed in this work (${\sim}1\%$).

\subsection{Third body in the spectra}
\label{sec:bf}

\begin{figure}
    \centering
    \includegraphics[trim = 7.5cm 7.5cm 3.5cm 1.5cm,clip,width=1.0\columnwidth]{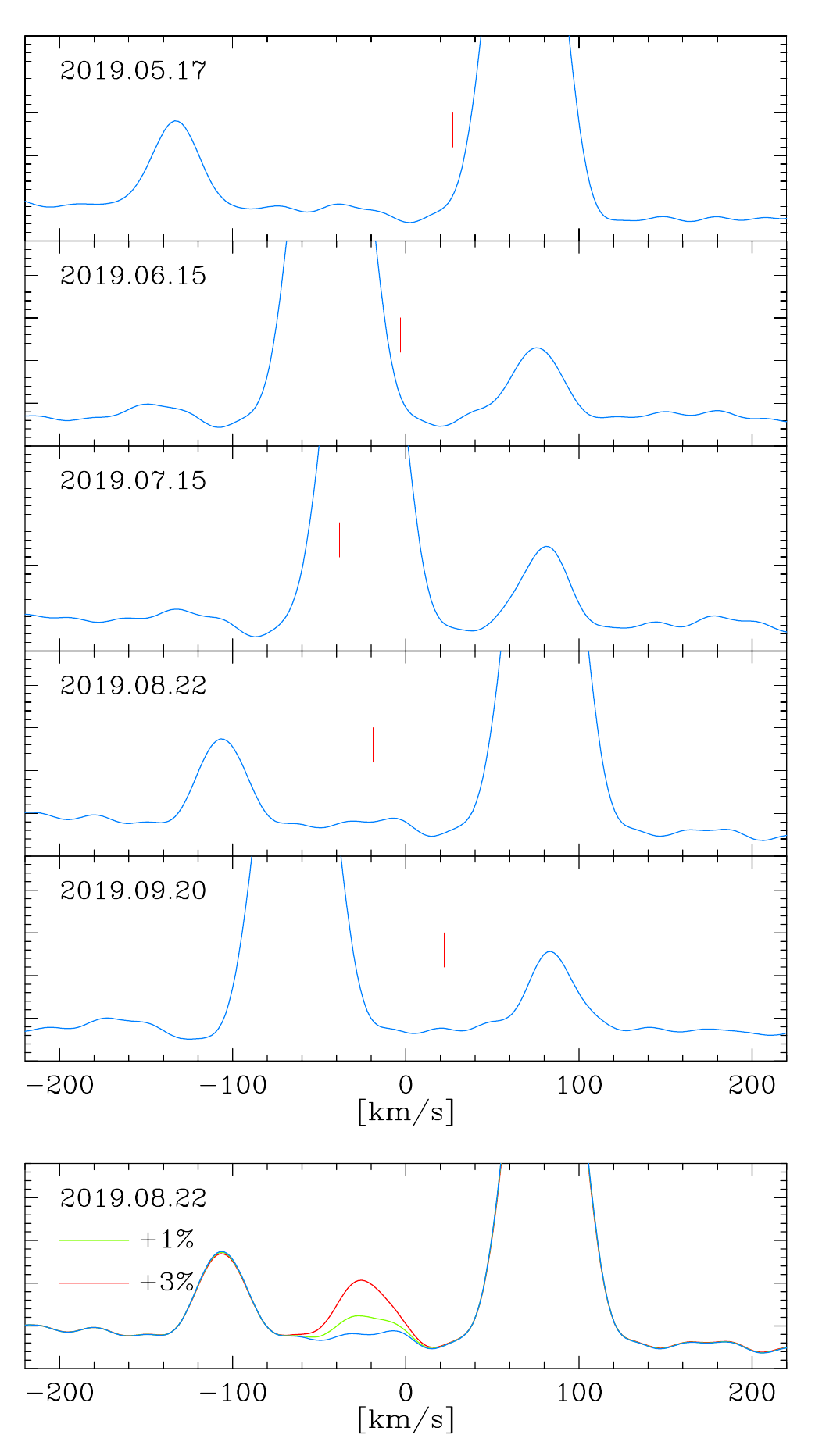}
    \caption{
    Broadening function (BF) analysis of CHIRON spectra. Upper panels show the BFs of five observations with the highest $S/N$, zoomed so the secondary peak and background are well visible (primary is out of scale). Red markers indicate velocities of a putative single star of mass equal to the minimum mass of the third body in the system, as found in the orbital solution. No prominent maxima are found on these positions. The lowest panel shows the BFs for one of the spectra, with artificially added third (single star) body that contributes about 1\% (green) and 3\% (red) to the total flux. Such contribution would have been detected.
    }
    \label{fig:bf}
\end{figure}

The high value of $M_B$, higher that $M_{Ab}$, suggests that the outer body should be easily detectable in the spectra. However, during the RV calculations we haven't noticed any prominent third peak in the cross-correlation functions, nor in the TODCOR maps. In order to verify this, we used the formalism of Broadening Function \citep[BF;][]{Rucinski1992,Rucinski2002}, and applied it to five CHIRON spectra that have the highest $S/N$. As a template we used a spectrum of $T_{\rm eff} = 5200$~K, $v_{\rm rot}=20$\,km\,s$^{-1}$ generated with ATLAS9. A single BF was generated for each of the echelle orders, and all the single-order BFs were then added in velocity domain, forming the final BF for a given observation. Additionally, we have calculated the expected RVs of the third star, if it had a mass of 0.871~$M_\odot$ (lower limit of mass, corresponding to the lowest flux contribution). Finally, to check if our approach would recover a small third-light flux, we took one of the spectra (from August 22, 2019) and injected artificial signals at the level of 1 and 3\% of the combined brightness of the inner binary. Results are shown in Figure~\ref{fig:bf}.

Two strong peaks, coming from the primary and secondary component, are clearly visible at positions corresponding to their RVs measured with TODCOR. However, no prominent third peak can be seen at positions expected for a 0.871~$M_\odot$ single star. One can also see that the 3\% additional signal is easily detectable, and the 1\% signal produces a distinctive peak as well. We therefore conclude that in our CHIRON spectra we would be able to detect the third light of at least 1\% level, and that a single star as massive as the secondary is not visible. 

Furthermore, at no other RV value we see any indication of a star as bright as the secondary, suggesting that the outer body, even if it is a binary, probably is not composed of such a star. Therefore, we can securely put a conservative {\it upper} limit to the total mass of the component B to be $2 \times M_{Ab} = 1.727$~M$_\odot$.

\section{Discussion}
\label{sec:discussion} 

\subsection{Physical parameters of \VCen{}}
\label{sec:param}

From our LC and RV analysis of \VCen{}, we determined the stellar masses and radii of the eclipsing pair with a better precision than in \citet{2015MNRAS.448.1937C}. In Table~\ref{tab:abs_param}, we presented the stellar parameters of each star Aa and Ab, along with their uncertainties, such as derived in this work. 

\begin{table}
\centering
\begin{minipage}{71mm}
	\caption{Stellar parameters and distance of \VCen{}.}
	\label{tab:abs_param}
	{
    \renewcommand{\arraystretch}{1.0}
	\begin{tabular}{@{}lccc@{}}
		\hline
		Parameter & Median & 84$\%$ interval & 16$\%$ interval \\
		\hline
		$a_{Aab}$ [R$_\odot$] & 10.121 & +0.037 & $-$0.038 \\
		$M_{Aa}$ [M$_\odot$] & 1.393 & +0.018 & $-$0.018 \\
		$M_{Ab}$ [M$_\odot$] & 0.863\,3 & +0.008\,1 & $-$0.008\,1 \\
		$R_{Aa}$ [R$_\odot$] & 1.407 & +0.014 & $-$0.014 \\
		$R_{Ab}$ [R$_\odot$] & 1.154 & +0.014 & $-$0.014 \\
		$\log g_{Aa}$ & 4.285\,5 & +0.008\,2 & $-$0.008\,1 \\
		$\log g_{Ab}$ & 4.249\,6 & +0.009\,9 & $-$0.009\,8 \\ \relax
		[Fe$/$H] & $-$0.18$^\textit{a}$ &  & \\
		$T_{{\rm eff},Aa}$ [K] & 6\,266$^\textit{a}$ & \multicolumn{2}{c}{$\pm$94} \\
		$T_{{\rm eff},Ab}$ [K] & 4\,650$^\textit{b}$ & \multicolumn{2}{c}{$\pm$900} \\
		$L_{Aa}$ [$\log\,(L$/L$_\odot)$] & 0.438 & +0.026 & $-$0.027 \\
		$L_{Ab}$ [$\log\,(L$/L$_\odot)$] & $-$0.25 & +0.30 & $-$0.36 \\
		$d$ [pc] & 95.7 & +5.8 & $-$4.3 \\
		$\pi$ [mas] & 10.45 & +0.50 & $-$0.60 \\
		\hline
	\end{tabular}
	}
    \textit{Notes}. $^\textit{a}$ From \citet{2009A&A...501..941H}.\\
    $^\textit{b}$ From temperature ratio obtained by \citet{2015MNRAS.448.1937C} using \PHOEBE{}.
\end{minipage}
\end{table}

As can be seen in Table~\ref{tab:abs_param}, we obtained values of stellar parameters that are in good agreement with those from \citet{2015MNRAS.448.1937C}, except for the radii of star Ab. Indeed, we found that $R_{Ab} = 1.154\,$R$_\odot$ instead of 1.10$\,$R$_\odot$, \ie a difference of 5$\%$. We confirm here the inflated radius of Ab, by about 52\% compared to its radius at the zero-age main sequence (ZAMS), which cannot be explained by activity alone and suggests that the star is in its PMS phase of evolution. In the following of the paper, this result will be compared with predictions of stellar models. Another difference comes from the precision of our derived stellar parameters. In particular, we reduced the uncertainties on the stellar mass and radius for the two stars to less than 1.3$\%$. We then adopted the same effective temperatures than \citet{2015MNRAS.448.1937C} to compute the intrinsic luminosities and the distance. In our calculations, we used the bolometric correction (BC) tables\footnote{\url{https://github.com/casaluca/bolometric-corrections}} from \citet{2018MNRAS.475.5023C,2018MNRAS.479L.102C}. By considering the values of $T_{\rm eff}$, $\log g$ and [Fe$/$H], listed in Table~\ref{tab:abs_param}, and an interstellar reddening $E(B - V) = 0$, we obtained ${\rm BC}_{Aa} = -0.017$ and ${\rm BC}_{Ab} = -0.451$. From the apparent visual magnitude of the system, $V_{\rm syst} = 8.415\,2(60)\,$mag, we then derived a photometric parallax of $10.45^{+0.50}_{-0.60}\,$mas for \VCen{}. This value can be directly compared with the trigonometric parallax from the \Gaia{} Data Release~2 (DR2; \citealt{2016A&A...595A...1G,2018A&A...616A...1G}), namely $\pi = 9.66 \pm 0.14\,$mas. We note that our parallax estimate does not match the new \Gaia{} DR2 value within their respective error bars. This difference can be explained by two factors. First, binaries and multiple stellar systems did not receive a special treatment during the \Gaia{} DR2 processing, \ie the sources were all treated as single stars. As a result, the parallax of such a multiple star can be affected by the orbital motion of the system \citep{2008IAUS..248...59P}. Second, our parallax estimate may be biased by the effective temperatures taken from \citet{2015MNRAS.448.1937C}. An accurate $T_{\rm eff}$ determination will then be performed in Section~\ref{sec:teff}.

Using the values of $M_{Aa}$ and $M_{Ab}$, associated with the quantity $f(M_B)$ in Table~\ref{tab:RV_param}, we also determined the mass of the third body. We found that $M_B = 0.871\pm0.020\,$M$_\odot$, which corresponds to a minimum value obtained by considering $i_{AB}=90^\circ$. As explained in Section~\ref{sec:bf}, we then attempted to search for signatures of the third body in CHIRON spectra with no success. A possible consequence is that the third body is itself a binary system with two low-mass stars of, for example, $0.45\,$M$_\odot$ each. From our analysis, we argue that \VCen{} is actually a quadruple star system with an outer period of 180.4~days instead of 351.5~days. Understanding the formation of close binaries in quadruple star systems represents a major issue in stellar astrophysics (see \citealt{2019MNRAS.482.2262H}, and references therein), which is beyond the scope of this paper.

\subsection{Kinematics}

In this work, we checked the validity of the Galactic space velocities ($U$, $V$, and $W$)\footnote{The values of $U$, $V$, and $W$ are positive in the directions of the Galactic centre, rotation, and north pole, respectively.} derived by \citet{2015MNRAS.448.1937C} for \VCen{}. To this end, we adopted the method developed by \citet{1987AJ.....93..864J} and implemented in the \IDL{} procedure \GALUVW{}\footnote{\url{https://idlastro.gsfc.nasa.gov/ftp/pro/astro/gal_uvw.pro}}. The input parameters of this procedure are the position ($\alpha$,$\delta$) at a reference epoch, the parallax $\pi$, the proper motion ($\mu_{\alpha\ast}$,$\mu_\delta$) and the systemic velocity $\gamma_{AB}$. We used the values provided by \citet{2018A&A...616A...1G}, except for $\gamma_{AB}$ where the value was taken from our best-fit RV solution in Table~\ref{tab:RV_param}. We then obtained $U = -24.11\pm0.39\,$km\,s$^{-1}$, $V = -32.87\pm0.49\,$km\,s$^{-1}$ and $W = -10.48\pm0.22\,$km\,s$^{-1}$. No correction for solar motion was applied. These results are in strong disagreement with those from \citet{2015MNRAS.448.1937C} that cannot be explained solely by the different values used as input parameters. Therefore, we suspect that their results are incorrect and that \VCen{} does not actually belong to the Hyades moving group. 

From our new values of the galactic velocities, we located \VCen{} at the edge of the Pleiades moving group (PMG; also called the Local Association) following the recent work of \citet{2017A&A...608A..73K}. The age of the PMG was estimated to be between 110--125$\,$Myr \citep{2018A&A...616A..10G}. It appears that the Pleiades age is more consistent with a young multiple star system than that of the Hyades ($\sim$625$\,$Myr; \citealt{1998A&A...331...81P}). However, we caution the reader that the 110--125$\,$Myr range is only a rough estimate of the systemic age. A more detailed analysis using stellar models is therefore required to precisely determine the individual ages of the eclipsing components (see Section~\ref{sec:models}).

\subsection{Effective temperature}
\label{sec:teff}

The goal of the present section is to constrain the effective temperature of the two eclipsing components with better precision than reported in previous studies. This parameter is fundamental in order to disentangle between different models for each star, and hence between different ages.

Here, we decided to use the procedure applied by \citet{1998A&A...330..600R} to a sample of detached double-lined eclipsing binaries belonging to the \Hipparcos{} catalogue. This procedure is based on the following expression:
\begin{equation}
T_{\rm eff}=\,T_{{\rm eff,}\odot} \left(10\,\pi\frac{R}{R_\odot}\right)^{-1/2}\!10^{-0.1(V+{\rm BC}-M_{{\rm bol,}\odot})},
\label{eq:teff}
\end{equation}
where the parallax $\pi$ is in arcsec and solar values are $T_{{\rm eff,}\odot}=5\,777\,$K and $M_{{\rm bol,}\odot}=4.74\,$mag. For each component, the stellar radius is taken from Table~\ref{tab:abs_param} and the apparent visual magnitude is computed from the values of $V_{\rm syst}$ and $l_2/l_1$ (the secondary-to-primary flux ratio). The corresponding value is $l_2/l_1=0.084\,3(44)$, which was derived using \JKTEBOP{}. We then obtained $V_{Aa}=8.503\,1(75)\,$mag and $V_{Ab}=11.190(53)\,$mag, with no correction for interstellar extinction. The effective temperature of the two stars was computed by adopting the Gaia DR2 parallax in equation~(\ref{eq:teff}). As explained in Section~\ref{sec:param}, the orbital motion of the system can affect the parallax measurement, although we expect this effect to be small. Finally, we adopted the BCs from \citet{2018MNRAS.475.5023C,2018MNRAS.479L.102C}, which depend on the effective temperature. We then proceeded in an iterative manner to compute the effective temperature of each star using equation~(\ref{eq:teff}). We started with a rough estimate of $T_{\rm eff}$ predicted by stellar models. This allowed us to determine a preliminary value of BC from the \citet{2018MNRAS.475.5023C,2018MNRAS.479L.102C} tables. The new $T_{\rm eff}$ value derived using equation~(\ref{eq:teff}), associated with the estimated BC, is then compared to the previous one. We repeated this procedure by adopting the new $T_{\rm eff}$ value to re-estimate the BC used in equation~(\ref{eq:teff}), until convergence. The final results are provided in Table~\ref{tab:teff}, and the corresponding BCs are found to be 0.001 and $-$0.576 for stars Aa and Ab, respectively.

\begin{table}
\centering
\begin{minipage}{47mm}
	\caption{Effective temperatures and luminosities of \VCen{} computed using the Gaia DR2 parallax.
	}
	\label{tab:teff}
	{
    \renewcommand{\arraystretch}{1.0}
	\begin{tabular}{@{}lcc@{}}
		\hline
		Parameter & Value & 1$\sigma$ error \\
		\hline
		$T_{{\rm eff},Aa}$ [K] & 6\,588 & 58 \\
		$T_{{\rm eff},Ab}$ [K] & 4\,475 & 68 \\
		$L_{Aa}$ [$\log\,(L$/L$_\odot)$] & 0.525 & 0.013 \\
		$L_{Ab}$ [$\log\,(L$/L$_\odot)$] & $-$0.319 & 0.025 \\
		\hline
	\end{tabular}
	}
\end{minipage}
\end{table}

\subsection{Comparison with stellar models}
\label{sec:models}

This section is dedicated to the comparison between the results from our LC and RV analysis of \VCen{} and the theoretical predictions from stellar models. The age determination of each of the two eclipsing stars will then help us to shed light on the evolutionary status of \VCen{}.

\subsubsection{MESA isochrones}

In order to determine the age of the two stars Aa and Ab, we generated a set of isochrones using a dedicated web interface\footnote{\url{http://waps.cfa.harvard.edu/MIST/}} based on the Modules for Experiments in Stellar Astrophysics (MESA; \citealt{2011ApJS..192....3P,2013ApJS..208....4P,2015ApJS..220...15P,2018ApJS..234...34P}) and developed as part of the MESA Isochrones and Stellar Tracks project (MIST v1.2; \citealt{2016ApJS..222....8D,2016ApJ...823..102C}). We considered in this work only the case of non-rotating stars ($v/v_{\rm crit} = 0$). For both stars, we first adopted the solar mixture from \citet{2009ARA&A..47..481A}, which corresponds to $Y_{\odot{\rm ,ini}}=0.270\,3$ and $Z_{\odot{\rm ,ini}}=0.014\,2$. We then searched for the isochrone that best matches the observed parameters ($R$, $M$, $T_{\rm eff}$) of each star. We adopted the effective temperatures derived in Section~\ref{sec:teff} and provided in Table~\ref{tab:teff}. Following the previous study of \citet{2015MNRAS.448.1937C}, we selected isochrones with ages lower than 30$\,$Myr. 

\begin{figure*}
	\includegraphics[trim = 0.5cm 0.0cm 1.0cm 1.0cm,clip,width=1.0\columnwidth,angle=0]{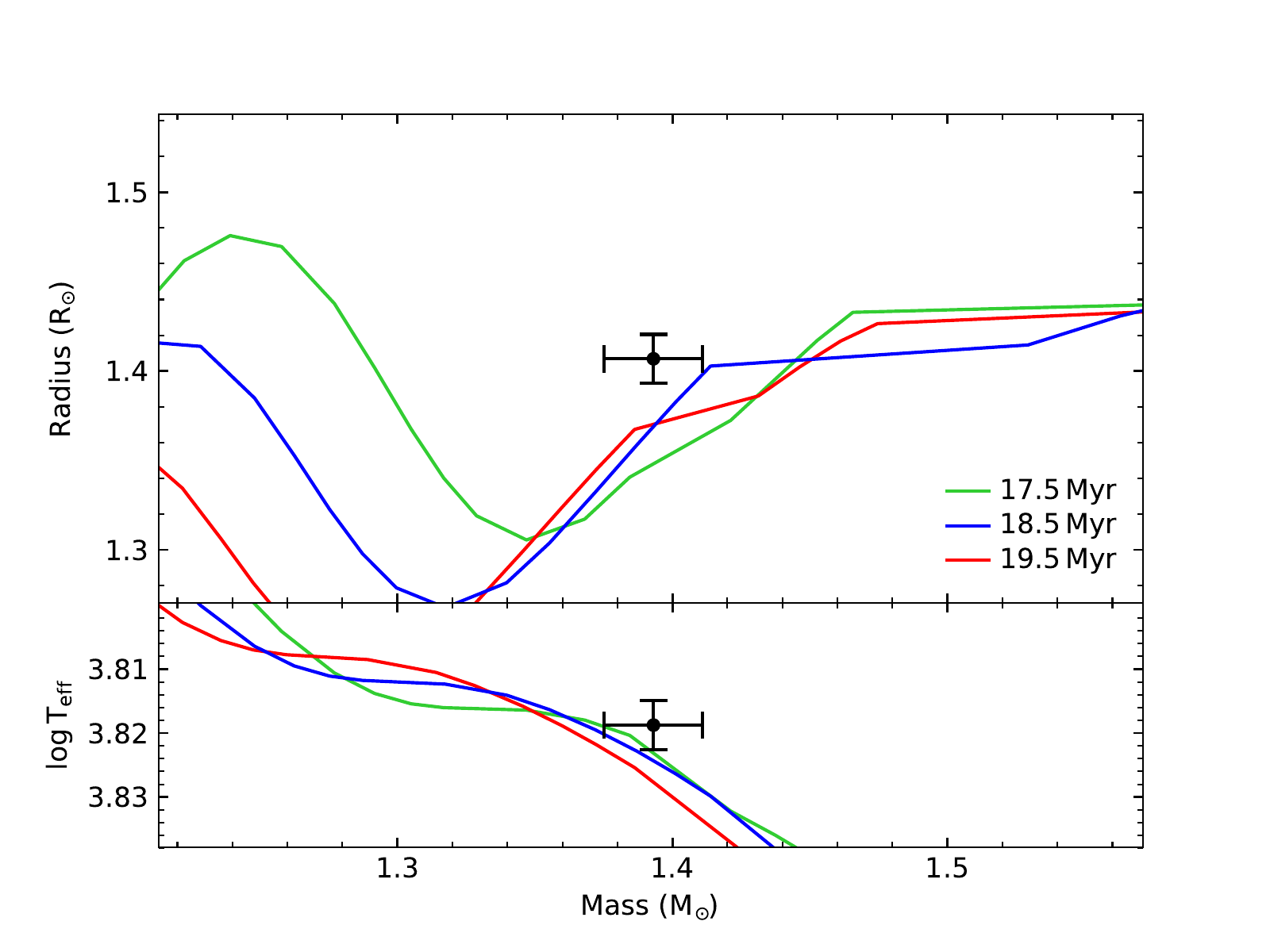}
	\includegraphics[trim = 0.5cm 0.0cm 1.0cm 1.0cm,clip,width=1.0\columnwidth,angle=0]{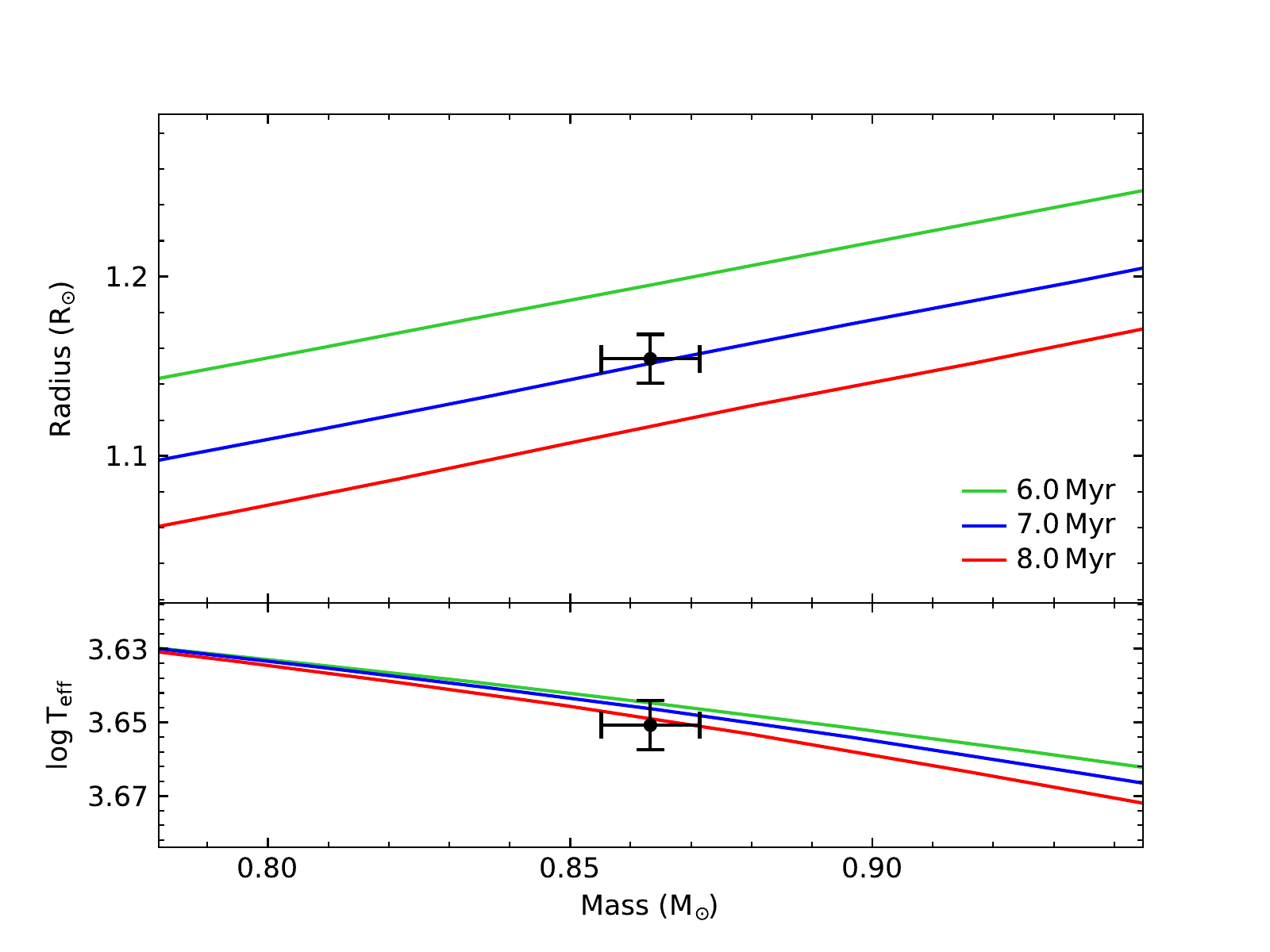}
    \caption{Comparison between the observed parameters of \VCen{} and the predictions from MESA isochrones. Left: Radius versus mass plane (upper panel) and $\log\,T_{\rm eff}$ versus mass plane (lower panel) for star Aa. Green, blue and red lines correspond to isochrones for ages of 17.5, 18.5 and 19.5 Myr, respectively. Black dots with error bars indicate the derived values of $R$, $M$ and $\log\,T_{\rm eff}$ with their corresponding 1$\sigma$ uncertainties. The $T_{\rm eff}$ value is taken from Table~\ref{tab:teff}. Right: Same as left, but for star Ab and isochrones with the ages indicated in the figure.}
    \label{fig:mesa_iso}
\end{figure*}

The comparison between the observed parameters from our analysis of \VCen{} and the predictions from MESA isochrones is shown in Fig.~\ref{fig:mesa_iso}. For star Aa, we found that the parameters $R$, $M$ and $T_{\rm eff}$ match well the 18.5-Myr isochrone within their 1$\sigma$ error bars, assuming a solar metallicity. For star Ab, we did not find an isochrone that simultaneously matches these parameters when assuming a solar metallicity. In particular, the predicted $T_{\rm eff}$ value is underestimated by about 400$\,$K for the best-matching isochrone. The effect of changing the metallicity was then investigated. Finally, we found an isochrone that matches the different parameters for an age of 7$\,$Myr and a metallicity of [Fe$/$H]$_{\rm ini}=-0.45$ (\ie $Y_{\rm ini}=0.256\,8$ and $Z_{\rm ini}=0.005\,2$), as shown in Fig.~\ref{fig:mesa_iso}. Furthermore, the precision on the derived parameters allowed us to distinguish between isochrones with an age difference of 1$\,$Myr. These results will be compared with those from another evolutionary code described in the next section.

\subsubsection{CESTAM stellar models}

We have also fitted the two stars using the CESTAM stellar evolution code \citep{2013A&A...549A..74M,2008Ap&SS.316...61M,Morel97}. Models were computed using the OPAL2005 equation of state \citep{2002ApJ...576.1064R} and the NACRE nuclear reaction rates \citep{ang99}. We have used the LUNA collaboration reaction rates for  $^{14}$N-burning \citep{Imbriani2005}. Convection is treated using the mixing-length theory \citep{1958ZA.....46..108B} with
a mixing length given by $\alpha H_P$. The value of $\alpha$ that calibrates a solar model computed without diffusion is $\alpha=1.64$. We have used the solar mixture of \citet{2009ARA&A..47..481A} in this work.
We adopted the OPAL opacity tables \citep{1996ApJ...464..943I} calculated with this solar mixture, complemented, at $T<10^4$ K, by the Wichita opacity data \citep{Ferguson2005}.
The atmosphere was treated using the Eddington grey approximation.

\begin{figure*}
	\includegraphics[trim = 0.0cm 0.0cm 0.5cm 1.3cm,clip,width=1.0\columnwidth,angle=0]{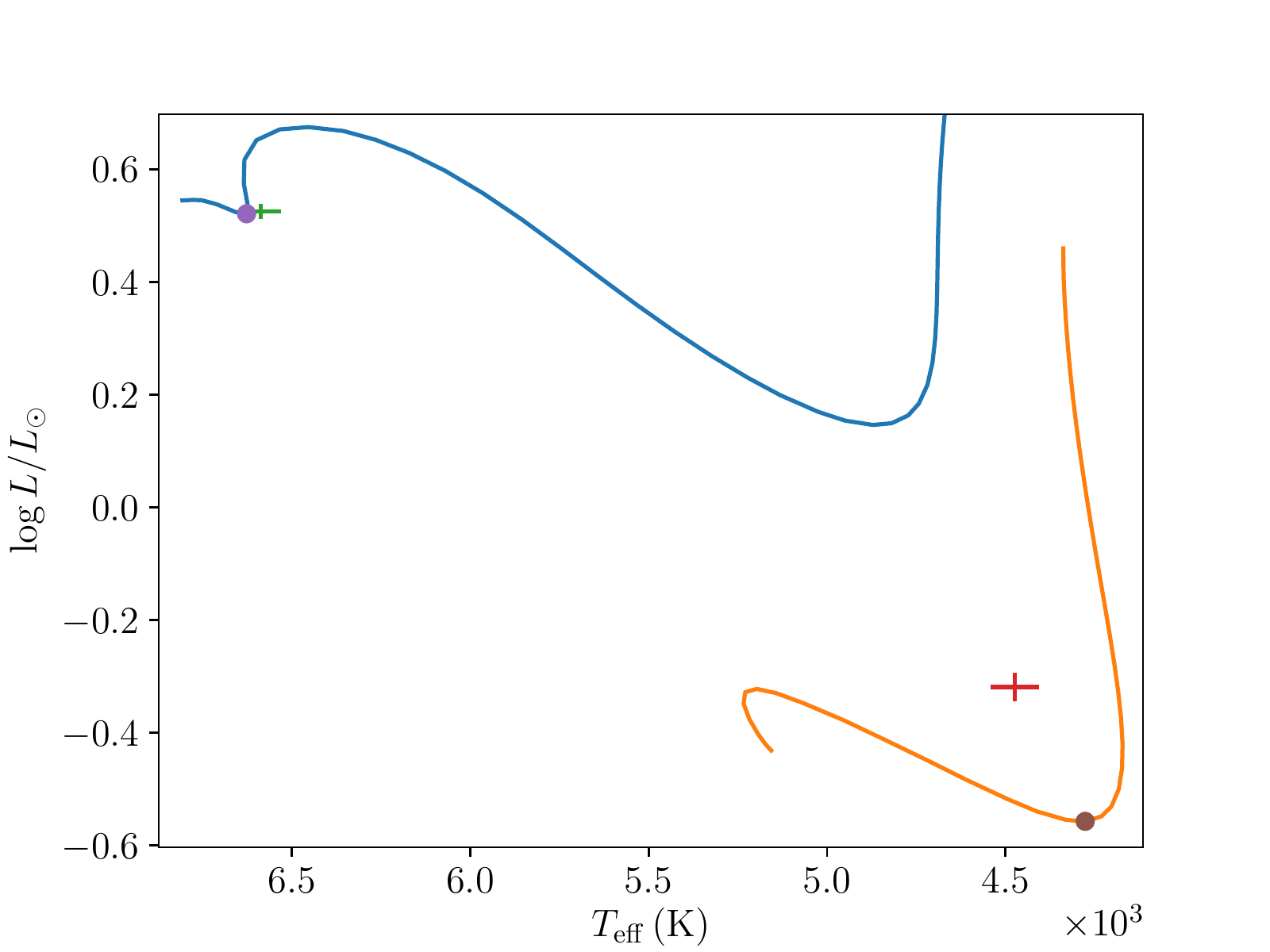}
	\includegraphics[trim = 0.0cm 0.0cm 0.5cm 1.3cm,clip,width=1.0\columnwidth,angle=0]{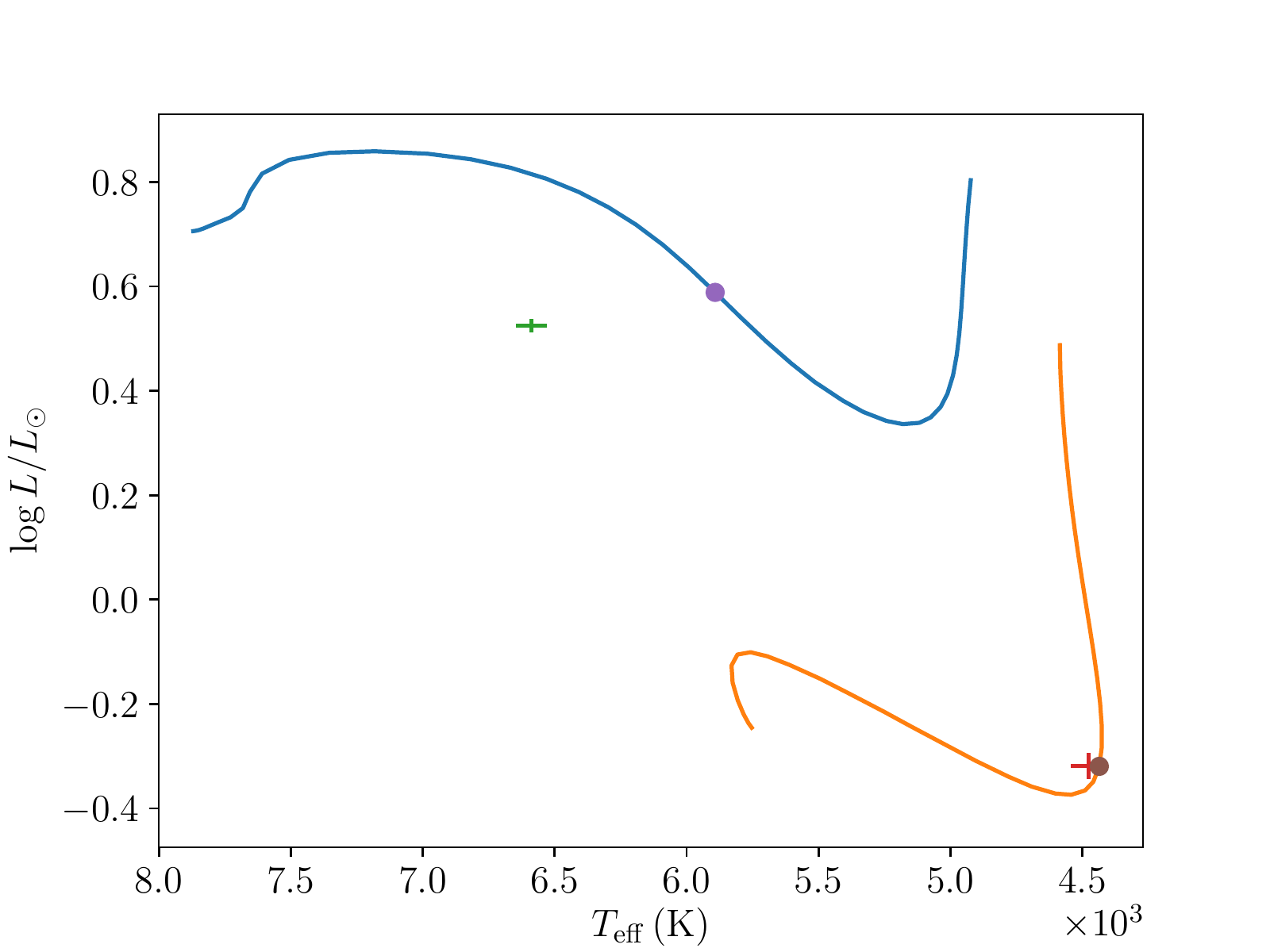}
    \caption{Evolutionary tracks in the HR diagram of models calculated with a solar metalicity (left panel) and $Z=0.005$ (right panel). Circles on the tracks indicate models with an age of 16 Myr (left panel) and 5.5 Myr (right panel).}
    \label{fig:HR_CESTAM}
\end{figure*}

We have found it impossible to fit both stars with the same age and chemical composition. The Aa component is 
easily fitted with a solar metalicity ($Z = 0.0134$, \citealt{2009ARA&A..47..481A}); its effective temperature and luminosity are reached at an age of 16 Myr. However, the model for the Ab component at the same age has an effective temperature and a luminosity that are lower than observed. This is shown in the left panel of Figure \ref{fig:HR_CESTAM}.

The Ab component can be fitted with a metalicity  $Z=0.005$. We obtain an effective temperature and a luminosity that are consistent with the observations at an age of $5.5 \pm 1$ Myr. However, the model for the Aa component is now too cold and bright at the same age, as seen in the right panel of Figure \ref{fig:HR_CESTAM}.

These results are consistent with the previous section.

\subsubsection{Evolutionary status of \VCen{}}

Using two different stellar models, we determined the individual age of each eclipsing component of the system from their observed parameters. We obtained a good agreement between MESA and CESTAM models. In particular, individual ages are found to be 16--18.5$\,$Myr and 5.5--7$\,$Myr for stars Aa and Ab, respectively. Furthermore, both models fail to reproduce the observed parameters of star Ab when assuming a solar metallicity. A lower metallicity is then required to properly fit the Ab component, while a solar metallicity is adopted for the Aa component.

Due to their common origin, stars belonging to a binary or multiple system are usually assumed to have the same age and initial chemical composition. However, the eclipsing components of \VCen{} appear to be non-coeval with a difference in age as high as $\sim$65\%. In addition to the age difference, the metallicity needed to fit the Ab component is lower by a factor of $\sim$2.7 than that of the Aa component, independently of the model used. If we consider that the secondary star has the same age and chemical composition as the primary star, then the observed parameters of the secondary are in clear disagreement with the predictions from both MESA and CESTAM. The secondary star is indeed hotter, larger and thus brighter than predicted by the evolutionary track for that age. In the opposite case, \ie when the age of the secondary star is considered, the effective temperature of the primary is too high by about 700$\,$K, making this hypothesis unlikely. Such discrepancies have already been reported by \citet{2016AJ....152....2L} for another PMS eclipsing binary, namely NP~Persei. In particular, the authors found that the two components of NP~Per cannot be fitted at a single age, implying a relative age difference of about 44\%. Stellar activity was proposed as a possible explanation for the discrepancies between the observed and predicted properties of NP~Per. However, based on the analysis of 13 PMS eclipsing binaries, \citet{2014NewAR..60....1S} concluded that activity alone cannot fully explain the discrepancies observed for this kind of systems. \citet{2014NewAR..60....1S} also noticed that half of their binaries have a tertiary companion. In the case of NP~Per, its short orbital period of 2.2~days could suggest the presence of an undetected companion in a wide orbit. 

The influence of a third body on the evolution of the eclipsing pair was investigated in detail by \citet{2014NewAR..60....1S}. Based on their conclusions, the evolutionary status of \VCen{} can be described as follows. This quadruple star system was likely formed 16--18.5$\,$Myr ago from a small gas cloud. The inner orbits were originally almost perpendicular to the outer orbit, allowing Lidov-Kozai cycles to take place. Both inner orbits acquired high eccentricity, thereby making the dynamical interaction between the four stars possible. Each sub-system is then circularised and tightened owing to their mutual influence. Once both inner orbits have been circularized, the two components of each system continue to interact by tidal effets, as their radii are still large compared to their separation. Such dynamical and tidal interactions may alter the stellar properties, resulting in the apparent non-coevality of the eclipsing components of \VCen{}.

\subsection{Confirmation of the quadruple nature of \VCen{}}

\subsubsection{Limitations of the single-star scenario for \VCen{}~B}
\label{sec:single_star}

If we consider that the third star has the same age as the primary, then the flux ratio between the tertiary and the secondary is about 0.6 in the \TESS{} band. In this case, the contribution of the third star to the total light is 7\% (more details will be given in Section~\ref{sec:third_light}). We can also consider that the third star has an inflated radius that is nearly equal to that of the secondary star, implying a similar luminosity at the age of the secondary isochrone. In addition, \citet{2017ApJ...844..103T} showed that the mutual inclination in compact low-mass triples is on average of $20^\circ$. Assuming this value, we obtain two possible configurations for the stellar system, which correspond to an inclination of the third-body orbit of about $61^\circ$ and $101^\circ$, respectively. The corresponding masses are respectively ${\sim}1.02\,$M$_{\odot}$ and ${\sim}0.89\,$M$_{\odot}$. The lower value is very close to that derived from the mass function. Depending on the isochrone, the flux ratio $l_3/l_2$ is expected to lie in the range 0.7--1.2. However, for the higher value of $M_B$, the flux ratio is found to be between 1.6 and 2.6. It appears that for a mass higher than ${\sim}0.95\,$M$_{\odot}$, the lines of the third star should be present in the spectrum. This limit corresponds to a reasonable mutual inclination of about $13^\circ$ ($i_{AB} \simeq 68^\circ$). Below this limit, the third light still represents more than 7\% of the total flux that is inconsistent with the results of the LC and BF analyses, where the third star contribution is not detected. Therefore, we think it is more likely that the third body is a binary system with two low-mass stars, each contributing to 1.5\% or less of the total flux when assuming stellar masses of 0.45$\,$M$_{\odot}$ (see Section~\ref{sec:third_light} below).

\subsubsection{Third-light contribution of the B sub-system}
\label{sec:third_light}

We investigated the impact on the derived stellar parameters of varying the third light during the light-curve analysis. For this, we adopted $l_3/l_{\rm tot} =$ $[0.03, 0.07, 0.12, 0.18]$, where the two first values are taken from the model predictions obtained in Section~\ref{sec:single_star}. The corresponding stellar parameters are listed in Table~\ref{tab:third_light}. 

It is notable that the secondary radius increases with increasing third light, whereas the primary radius decreases. The last case in Table~\ref{tab:third_light} corresponds to a third star having almost the same flux contribution as the secondary ($l_2/l_{\rm tot} = 21\%$). In this case, the primary and secondary stars have similar radii of 1.250$\,$R$_\odot$ and 1.245$\,$R$_\odot$, respectively, while their masses remain nearly unchanged compared to our reference model. The consequence is that the primary star appears to be older and the secondary younger than previously estimated, implying an even higher age difference. In addition, when we consider the case of a second binary that contributes to 3\% of the total flux, the primary and secondary radii differ by $-1.5\sigma$ and $+0.3\sigma$ from the reference values, respectively. These values are in better agreement than those obtained considering a single star, which contributes 7\% to the total flux. In this latter case, the stellar radii differ by $-4.2\sigma$ and $+1.5\sigma$ for the primary and the secondary, respectively. For all cases, the goodness of fit is similar due to the correlations linking the third-light
parameter to the other parameters ($i$, $r_1$ and $r_2$).

\begin{table}
\centering
\begin{minipage}{62mm}
    \caption{
    Stellar parameters of \VCen{} as a function of the third-light contribution. The first line corresponds to our reference model (see Table~\ref{tab:abs_param}).
    }
    \label{tab:third_light}
    {
    \renewcommand{\arraystretch}{1.0}
    \begin{tabular}{@{}cccccc@{}} 
        \hline
        $l_3/l_{\rm tot}$ & $i_A$ & $R_{Aa}$ & $R_{Ab}$ & $M_{Aa}$ & $M_{Ab}$ \\ \relax
        [\%] & [$^\circ$] & [R$_\odot$] & [R$_\odot$] & [M$_\odot$] & [M$_\odot$] \\
        \hline
        0 & 81.38 & 1.407 & 1.154 & 1.393 & 0.863\,3 \\
		3 & 81.49 & 1.387 & 1.158 & 1.392 & 0.862\,5 \\
		7 & 81.66 & 1.350 & 1.175 & 1.390 & 0.861\,4 \\ 
		12 & 81.77 & 1.314 & 1.198 & 1.389 & 0.860\,7 \\
		18 & 81.92 & 1.250 & 1.245 & 1.387 & 0.859\,7 \\
        \hline
    \end{tabular}
    }
\end{minipage}
\end{table}


\section{Summary}
\label{sec:summary}

The aim of the present work was to perform a new analysis of \VCentauri{}, a multiple star system that contains a close eclipsing binary. For this, we made use of the most recent observations of the system from the \Solaris{} network, the \TESS{} space telescope, and the CHIRON spectrograph. The combined analysis of the light curves and radial velocity measurements allowed us to derive the mass and radius of each eclipsing component with a precision better than 1.3\%. The resulting values for the primary component are $M_{Aa} = 1.393\pm0.018\,$M$_\odot$ and $R_{Aa} = 1.407\pm0.014\,$R$_\odot$. For the secondary, we found that $M_{Ab} = 0.863\,3\pm0.008\,1\,$M$_\odot$ and $R_{Ab} = 1.154\pm0.014\,$R$_\odot$, where the inflated radius confirms the PMS nature of the system. We also confirmed the 2.5-day orbital period of the eclipsing pair, whereas the eccentricity was found to be slightly different from zero ($e=0.01$). However, regarding the outer orbit, we obtained significantly different results than those reported in the literature. Thanks to the additional measurements from CHIRON, we derived a new orbital solution assuming an outer period of 180.4 days, instead of 351.5 days, and a minimum mass for the third body of $0.871\pm0.020\,$M$_\odot$. A consequence of this result is that the third body is actually a binary system with two low-mass stars that are not detectable from our observations. \VCentauri{} is thus a quadruple star system consisting of two close pairs orbiting each other with a 180-day period.

Finally, we compared the observed parameters of each eclipsing components with the predictions from two independent stellar evolution codes, namely MESA and CESTAM. In addition to the mass and radius, we also used the effective temperatures derived in this work to better constrain the individual ages of the eclipsing components. From their radii and apparent magnitudes, we found the effective temperatures of the stars to be $T_{{\rm eff},Aa} = 6\,588\pm58\,$K and $T_{{\rm eff},Ab} = 4\,475\pm68\,$K when adopting the \Gaia{} DR2 parallax. We then obtained ages of 16--18.5$\,$Myr and 5.5--7$\,$Myr for stars Aa and Ab, respectively. Despite the good agreement between MESA and CESTAM models, we failed to reproduce the observed parameters by assuming the same age and chemical composition for both stars. In particular, it is noticeable that the secondary star appears both larger and hotter than predicted at the age of the primary. For \VCen{}, the relative age difference is particularly high ($\sim$65\%). However, it is likely that the stars in such a close quadruple system experienced strong dynamical and tidal interactions, possibly affecting the observed stellar parameters. In conclusion, the case of \VCentauri{} provides a real challenge for theoreticians to model PMS stars in multiple systems, and to account for their apparent non-coevality.


\section*{Acknowledgements}

Based on data collected with Solaris network of telescopes of the Nicolaus Copernicus Astronomical Center of the Polish Academy of Sciences.
This paper includes data collected by the \TESS{} mission, which are publicly available from the Mikulski Archive for Space Telescopes (MAST). Funding for the \TESS{} mission is provided by NASA's Science Mission directorate. This research made use of Photutils, an Astropy package for detection and photometry of astronomical sources \citep{Bradley_2019_2533376}. This work has made use of data from the European Space Agency (ESA) mission \Gaia{} (\url{https://www.cosmos.esa.int/gaia}), processed by the \Gaia{} Data Processing and Analysis Consortium (DPAC, \url{https://www.cosmos.esa.int/web/gaia/dpac/consortium}). Funding for the DPAC has been provided by national institutions, in particular the institutions participating in the \Gaia{} Multilateral Agreement. This research has made use of NASA's Astrophysics Data System Bibliographic Services, the SIMBAD database, operated at CDS, Strasbourg, France and the VizieR catalogue access tool, CDS, Strasbourg, France. The original description of the VizieR service was published in \citet{2000A&AS..143...23O}.

We acknowledge support provided by the Polish National Science Center (NCN) through grants 2017/27/B/ST9/02727 (FM, MK), 2016/21/B/ST9/01613 (KGH) and 2015/16/S/ST9/00461 (MR).

Finally, we also thank the anonymous referee for comments that helped to improve this paper.


\section*{Data availability}

The data underlying this article will be shared on reasonable request to the corresponding author.




\bibliographystyle{mnras}
\bibliography{fredm} 



\appendix

\section{Radial velocities}

In Table~\ref{tab:RV_obs}, we listed all the RV measurements of \VCen{} used in this study, together with the final measurement errors $\sigma$. For the sake of clarity, we kept the notation introduced by \citet{2015MNRAS.448.1937C}, where indices 1 and 2 refer respectively to the primary and secondary components (Aa and Ab) of the eclipsing pair. The last column shows the telescope/spectrograph used, coded as follows: 5/P\,=\,OUC 50-cm/PUCHEROS, E/C\,=\,Euler 1.2-m/CORALIE, C/C\,=\,CTIO 1.5-m/CHIRON.

\begin{table*}
\centering
\begin{minipage}{78mm}
	\caption{Individual RV measurements of \VCen{} used in this work. All values are given in km\,s$^{-1}$.}
	\label{tab:RV_obs}
	{
    \renewcommand{\arraystretch}{1.0}
	\begin{tabular}{@{}lrrrrc@{}}
		\hline
		JD$-2\,450\,000$ & \thead{$v_1$} & \thead{$\sigma_1$} & \thead{$v_2$} & \thead{$\sigma_2$} & Tel./Sp. \\
		\hline
        5\,714.615\,861 & 45.958    & 0.646 &         -- &    -- & 5/P \\
        5\,736.539\,995 & 64.395    & 0.494 & $-$127.519 & 2.640 & 5/P \\
        5\,737.639\,889 & $-$67.029 & 0.523 & 88.377     & 4.198 & 5/P \\
        5\,750.604\,835 & $-$62.826 & 2.446 &         -- &    -- & 5/P \\
        5\,751.584\,224 & 74.651    & 0.536 & $-$126.370 & 4.324 & 5/P \\
        6\,066.642\,808 & 47.460    & 1.498 &         -- &    -- & 5/P \\
        6\,066.665\,643 & 51.655    & 0.841 &         -- &    -- & 5/P \\
        6\,078.565\,477 & $-$36.335 & 2.129 &         -- &    -- & 5/P \\
        6\,080.625\,298 & $-$89.867 & 0.163 & 112.325    & 1.503 & E/C \\
        6\,081.564\,728 & 52.113    & 0.228 & $-$116.745 & 1.075 & E/C \\
        6\,179.474\,281 & $-$26.024 & 0.167 & 80.336     & 0.885 & E/C \\
        6\,346.690\,592 & $-$12.831 & 0.169 & 67.855     & 0.876 & E/C \\
        6\,348.857\,536 & $-$55.020 & 0.165 & 136.192    & 1.064 & E/C \\
        6\,349.894\,755 & 94.865    & 0.194 & $-$107.687 & 1.017 & E/C \\
        6\,397.520\,928 & 38.353    & 0.112 & $-$71.655  & 0.772 & E/C \\
        6\,398.517\,694 & $-$77.575 & 0.116 & 112.000    & 0.951 & E/C \\
        6\,497.610\,599 & $-$67.667 & 0.157 & 133.439    & 0.797 & E/C \\
        6\,498.610\,654 & 64.361    & 0.113 & $-$78.099  & 0.942 & E/C \\
        8\,621.784\,282 & 69.037    & 0.134 & $-$134.418 & 0.739 & C/C \\
        8\,650.705\,804 & $-$42.656 & 0.146 & 76.048     & 0.869 & C/C \\
        8\,680.534\,650 & $-$23.882 & 0.134 & 81.379     & 0.520 & C/C \\
        8\,696.577\,977 & 73.578    & 0.337 & $-$70.828  & 2.379 & C/C \\
        8\,718.466\,921 & 80.631    & 0.131 & $-$106.965 & 1.035 & C/C \\
        8\,747.475\,367 & $-$62.708 & 0.138 & 82.865     & 0.655 & C/C \\
		\hline
	\end{tabular}
    }
\end{minipage}
\end{table*}


\bsp	
\label{lastpage}
\end{document}